\documentclass[%
preprint,
showpacs,preprintnumbers,
 amsmath,amssymb,
 aps,
pre,
]{revtex4-1}
\pdfoutput=1
\usepackage{graphicx}
\usepackage{dcolumn}
\usepackage{bm}
\usepackage{multirow}

\usepackage[colorlinks,linkcolor=blue,citecolor=blue]{hyperref}
\usepackage{subfigure}
\usepackage{color}
\usepackage{url}
\usepackage{hyperref}

\usepackage{amsmath,amssymb}
\usepackage{subfigure}
\usepackage{caption}    
\usepackage{color}
\usepackage{url}
\usepackage{lineno,hyperref}
\usepackage{graphicx}
\usepackage{dcolumn}
\usepackage{bm}
\modulolinenumbers[5]

\begin{document}




\title{An Implicit Gas-Kinetic Scheme for Turbulent Flow on Unstructured Hybrid Mesh}

\author{Dongxin Pan}
\email{chungou@mail.nwpu.edu.cn}
\author{Chengwen Zhong}%
\email{Corresponding author: zhongcw@nwpu.edu.cn}
\author{Congshan Zhuo}%
\email{zhuocs@nwpu.edu.cn}

\affiliation{National Key Laboratory of Science and Technology on Aerodynamic Design and Research, Northwestern Polytechnical University, Xi'an, Shaanxi 710072, China.}

\date{\today}

\begin{abstract}
In this study, an implicit scheme for the gas-kinetic scheme (GKS) on the unstructured hybrid mesh is proposed. The Spalart-Allmaras (SA) one equation turbulence model is incorporated into the implicit gas-kinetic scheme (IGKS) to predict the effects of turbulence. The implicit macroscopic governing equations are constructed and solved by the matrix-free lower-upper symmetric-Gauss-Seidel (LU-SGS) method. To reduce the number of cells and computational cost, the hybrid mesh is applied. A modified non-manifold hybrid mesh data(NHMD) is used for  both unstructured hybrid mesh and uniform grid. Numerical investigations are performed on different 2D laminar and turbulent flows. The convergence property and the computational efficiency of the present IGKS method are investigated. Much better performance are obtained compared with the standard explicit gas-kinetic scheme. Also, our numerical results are found to be in good agreement with experiment data and other numerical solutions, demonstrating the good applicability and high efficiency of the present IGKS for the simulations of laminar and turbulent flows.
\end{abstract}


\maketitle

\section{Introduction}
\label{Introduction}

Over the past half-century, the gas-kinetic theory has been extensively studied for incompressible and compressible fluid flows. As an alternative method based on the Bhatnagar-Gross-Krook (BGK) collision model, the gas kinetic scheme (GKS) proposed by Xu and Prendergast~\cite{xu1994numerical} has been rapidly developed in the last two decades and has attracted an increasing amount of attention from the CFD community. Through new modifications as reported in the literature~\cite{xu2001gas}, the revised GKS scheme is of a certain advantage of resolving dissipative structure, especially in capturing shock waves. In the traditional Navier-Stokes (NS) solvers, the gas is assumed to stay in two equilibrium states on both sides of the shock wave and the shock wave appears as a discontinuity; that is to say, the physical dissipation in a cell size is replaced by numerical one. In the flux reconstruction approach, an additional artificial dissipation is also introduced by NS solver due to application of upwind scheme and/or central difference method. To remove the spurious dissipation, a general non-equilibrium state is considered for constructing the initial gas distribution function at each time step and the equilibrium state at a cell interface utilizing the Chapman-Enskog expansion~\cite{xu2001gas}. In the complex flow simulations, the GKS can give us more real description from the aspect of physical evolution than traditional NS solvers.

In practice, as reported by May et al.~\cite{may2007improved}, the flux evaluation in GKS is slightly more complicated compared with that generally used in the other finite volume method (FVM) solvers, which means that GKS takes more computational cost than other FVM solvers under the same situation. So, in practical applications, acceleration methods for GKS must be developed. To reduce the computational time, the parallel implementation is a good strategy. Ilgaz and Tuncer~\cite{ilgaz2006parallel} investigated the parallel implementation of the gas-kinetic BGK scheme on unstructured grids by using the domain decomposition method. Kumar et al.~\cite{kumar2013weno} proposed a parallelized WENO-enhanced GKS for the three-dimensional (3D) direct simulations of compressible transitional and turbulent flows.

On the other hand, for the steady flow simulations concerned in this study, using implicit scheme is an efficient way to accelerate convergency and reduce computational cost. Chit et al.~\cite{chit2004implicit} applied an approximate factorization and alternating direction-implicit (AF-ADI) method for the GKS simulations of the inviscid compressible flow on structured grids, in which an improved algorithm was achieved with fast convergence and the capability for using a large Courant-Friedrichs-Lewy (CFL) number, and this implicit method exhibited better results than the traditional explicit method. Recently, Li et al.~\cite{li2014implicit} presented an implicit gas-kinetic method with the matrix-free lower-upper Symmetric Gauss-Seidel (LU-SGS) time marching implicit scheme to simulate the hypersonic inviscid flows on unstructured mesh, in which the implicit CFL number was determined by decreasing residual and the good robustness of this kind of implicit gas-kinetic scheme (IGKS) was validated in their works. However, the convergent behavior was not shown with the residual curve in their works. As mentioned above, several kinds of IGKS have been applied for compressible flow simulations~\cite{chit2004implicit,li2014implicit}. For unsteady flows, a dual time-stepping strategy of gas-kinetic scheme is proposed by Li et al~\cite{PhysRevE.95.053307}. In their works, fluid flows from laminar to turbulent and from incompressible to compressible are accurately simulated with better efficiency than previous method. However, there were few studies on implementing IGKS on unstructured hybrid grids for the compressible turbulent flows around/in complex geometries. This study focus on presenting an IGKS with the LU-SGS scheme on unstructured hybrid grids for complex turbulent flows, and the convergent behavior of IGKS is also investigated in detail.

For the simulations of compressible flow around/in complex geometries, the unstructured hybrid grid becomes a promising choice, due to its advantage in balancing the accuracy and the computational cost. For example, for flow around complex realistic configuration, the body fitted grid can be used to resolve the boundary layer region, while the unstructured grid with a suitable growth rate can be applied to fill all other computational domain. Besides, the grid refinement and coarsening are quite easy to be implemented~\cite{mavriplis1997unstructured}, compared to the structured or block-structured grid technique which relies on regular connectivity of quadrilateral or hexahedral cells. In problems with complex configurations, high-order finite volume method under unstructured grids can also obtain more elaborate and precise results~\cite{liu2016high}. In compressible cases, limiter for unstructured grid is easy to implement~\cite{liu2017accuracy}. As we known, the complexity of hybrid grids, including elements, edges, nodes, and connectivity, needs an efficient mesh data structure to reduce the extra computational costs. In this study, the extended non-manifold hybrid mesh data (NHMD)~\cite{ebeida2009mesh,FLD4239} is employed to build an accessible library for mesh.

Most flows encountered in engineering applications are of turbulent nature. For the compressible turbulent flow simulations, compared to the direct numerical simulation (DNS) and the large eddy simulation (LES), the turbulence models require coarse grid resolution and have been validated to be suitable for the turbulent flow simulations in engineering. In particular, the one-equation Splalart-Allmaras (SA) model~\cite{spalart1992one} became quite popular because of its satisfactory results for a wide range of flow problems and its reliable numerical properties. On the basis of the explicit GKS, we have successfully carried out the simulation of compressible turbulent flows with shock waves on unstructured meshes, in which the SA turbulence model {\color{red}{~\cite{FLD4239}}} and SST turbulence model {\color{red}{~\cite{PhysRevE.95.053307}}} are incorporated to include the effect of turbulence. In this study, the SA turbulence model is also selected on account of its good accurate and stability under relatively coarse grid near the wall, as well as its low computational cost. Other choices, such as the two-equation models, might also be good candidates, which will be developed in the future.

In this paper, an IGKS coupled with SA turbulence model is proposed for the incompressible and compressible flow simulations on unstructured hybrid mesh, and several flow simulations including the lid-driven cavity flow, the laminar and the turbulent flow around a flat plate, as well as the turbulent flow around a multi-element airfoil are performed here.

The rest of the paper is organized as follows. The basic GKS proposed by Xu and Prendergast~\cite{xu1994numerical}, the matrix-free LU-SGS scheme and the implementation of IGKS, the expanded NHMD structure and the coupled SA turbulence model, as well as four different boundary conditions are described in Section \ref{sec:IGKS}. Then, the numerical validations and several turbulent flow cases are carried out to show the accuracy and reliability of the present IGKS-NHMD method in Section \ref{sec:NUMERICAL RESULTS}. Finally, some remarks concluded from this study are grouped in Section \ref{sec:conlustion}.

\section{IMPLICIT METHOD for GAS-KINETIC SCHEME}\label{sec:IGKS}
\subsection{GAS-KINETIC SCHEME}\label{sec:GKS}
The two-dimensional (2D) gas-kinetic scheme based on BGK model is written as~\cite{xu1994numerical}
\begin{equation}\label{Eq01}
\frac{{\partial f}}{{\partial t}} + u\frac{{\partial f}}{{\partial x}} + v\frac{{\partial f}}{{\partial y}} = \frac{{g - f}}{\tau },
\end{equation}
where $f$ and $g$ are the gas distribution functions, which are the functions of space $(x, y)$, particle velocity ${\vec u} = (u,v)$, time $t$, and internal variable $\xi$. $\tau$ is the particle collision time. $g$ is the Maxwell distribution function which has the following form
\begin{equation}\label{Eq02}
g = \rho {\left( {\frac{\lambda }{\pi }} \right)^{\frac{{K + 2}}{2}}}{e^{ - \lambda \left( {{{\left( {u - U} \right)}^2} + {{\left( {v - V} \right)}^2} + {\xi ^2}} \right)}},
\end{equation}
where $K$ is the internal freedom degree with $K = 3$ for the 2D diatomic molecule gas flows. The variable $\lambda = m/(2RT)$, $m$ is the molecular mass, $R$ is the Boltzmann constant, and $T$ is the temperature. $\rho$ is the density, $U$ and $V$ are the $x$ and $y$ components of the macroscopic velocity in 2D, respectively. Note that, for the gas system with K freedom degree, the square of internal variable $\xi^2$ can be taken as
\begin{equation}
{\xi^2}={\xi_1^2}+{\xi_2^2}+...+{\xi_K^2} .
\end{equation}

On the cell interface, the general solution $f$ of Eq.~(\ref{Eq01}) at time $t$ can be given as~\cite{xu1994numerical,xu2001gas}:
\begin{equation}\label{Eq03}
f\left( {{\vec x}_{j + \frac{1}{2}},t,{\vec u},\xi } \right) = \frac{1}{\tau }\int_0^t {g\left( {\vec x'},t',{\vec u},\xi \right){e^{ - \frac{{\left( {t - t'} \right)}}{\tau }}}dt'}  + {e^{ - \frac{t}{\tau }}}{f_0}\left( {{{\vec x }_{j + \frac{1}{2}}} - {\vec u}  t}, {\vec u}, \xi \right) ,
\end{equation}
where the particle trajectory $\vec x' = {\vec x }_{j + \frac{1}{2}} -\vec u (t-t')$. For simplicity, setting ${\vec x} _{j + \frac{1}{2}} = \left( {0,0} \right)$ t the cell interface in the following text.

To include the physical nature of non-equilibrium phenomena in the complex flows, the deviation of a distribution function away from a Maxwellian velocity distribution should be applied to construct the gas distribution function. In Eq.~(\ref{Eq03}) the initial gas distribution function $f_0$ at the beginning of each time step is written as
\begin{equation}\label{Eq04}
{f_0} = \left\{ \begin{array}{l}
{g^l}\left( {1 + a_n^l x + a_t^l y - \tau \left( {a_n^lu + a_t^lv + {A^l}} \right)} \right) \quad x \le 0,\\
{g^r}\left( {1 + a_n^r x + a_t^r y - \tau \left( {a_n^ru + a_t^rv + {A^r}} \right)} \right) \quad x > 0,
\end{array} \right.
\end{equation}
where $g^l$ and $g^r$ represent the Maxwellian distribution functions on the left and right sides of interface, $a_n^l$, $a_n^r$, $a_t^l$, $a_t^r$, $A^l$ and $A^r$ are related to the derivatives of the Maxwellian velocity distribution in space and time. The additional time and the spatial derivative terms in Eq.~(\ref{Eq04}), $- \tau \left( {a_n^lu + a_t^lv + {A^l}} \right)$ and $- \tau \left( {a_n^ru + a_t^rv + {A^r}} \right)$, denote the non-equilibrium effects, which have no direct contribution to the macroscopic conservative variables.

All the derivative terms in Eq.(~\ref{Eq04}), including $a_n^l$, $a_t^l$, $a_n^r$, $a_t^r$, $A^l$, and $A^r$ have the form of Taylor expansion
\begin{equation}\label{100}
\begin{aligned}
a &= a_1 + a_2 u + a_3 v + \frac{1}{2} a_4 \left( u^2 + v^2 + {\xi ^2} \right),\\
A &= A_1 + A_2 u + A_3 v + \frac{1}{2} A_4 \left( u^2 + v^2 + {\xi ^2} \right).
\end{aligned}
\end{equation}
These terms can be obtained from the relations between the gas distribution functions and the macroscopic variables.
\begin{equation}\label{101}
\begin{aligned}
\int {ga_n \bm{\psi} d\Xi } &= \frac{\partial \bm{W}}{\partial x} ,\\
\int {ga_t \bm{\psi} d\Xi } &= \frac{\partial \bm{W}}{\partial y} ,
\end{aligned}
\end{equation}
where $d\Xi = dudvd\xi$, $\bm{\psi} = \left(1,u,v,({{u^2} + {v^2} + {\xi ^2}})/2\right)^{\text{T}}$ is the vector of collisional invariants, and the macroscopic variables $\bm{W}=(\rho,\rho U,\rho V,\rho E)^{\text{T}}$.

To reach second-order accuracy, the equilibrium distribution $g$ should contain the first-order spatial and time derivative terms,
\begin{equation}\label{Eq05}
g = \left\{ \begin{array}{l}
{g_0}\left( {1 + \bar a_n^lx + {{\bar a}_t}y + \bar At} \right) \quad x \le 0,\\
{g_0}\left( {1 + \bar a_n^rx + {{\bar a}_t}y + \bar At} \right) \quad x > 0,
\end{array} \right.
\end{equation}
where ${\bar a}_n^l$, ${\bar a}_n^r$, ${\bar a}_t$, and $\bar A$ are spatial derivatives and time derivative of distribution function for equilibrium state. In Eq.~(\ref{Eq04}) and Eq.~(\ref{Eq05}), $l$, $r$, and $0$ represent left, right of the interface and on the interface, respectively.

In the present simulations on the unstructured mesh, all spatial derivative terms in Eq.~(\ref{Eq04}) can be obtained by the least squares method. Taking the simple schematic grid shown in Fig.~\ref{fig:Fig01} for example, firstly, the fitting formulation is applied
\begin{equation}\label{Eq06}
\bm{W} = {\bm{W}_0} + {\bm{b}_x}\left( {x - {x_0}} \right) + {\bm{b}_y}\left( {y - {y_0}} \right).
\end{equation}
To determine the derivative terms $\bm{b}_x$ and $\bm{b}_y$, the least-squares regression equations can be reconstructed as
\begin{equation}\label{Eq07}
\sum\limits_{k = 1}^N {\left[ \begin{array}{c}
{\left( {{x_k} - {x_0}} \right)^2}\\
\left( {{x_k} - {x_0}} \right)\left( {{y_k} - {y_0}} \right)
\end{array} \right.} \left. \begin{array}{c}
\left( {{x_k} - {x_0}} \right)\left( {{y_k} - {y_0}} \right)\\
{\left( {{y_k} - {y_0}} \right)^2}
\end{array} \right]\left[ \begin{array}{l}
{\bm{b}_x}\\
{\bm{b}_y}
\end{array} \right] = \sum\limits_{k = 1}^N {\left[ \begin{array}{l}
\left( {{x_k} - {x_0}} \right)\left( {{\bm{W}_k} - {\bm{W}_0}} \right)\\
\left( {{y_k} - {y_0}} \right)\left( {{\bm{W}_k} - {\bm{W}_0}} \right)
\end{array} \right]} ,
\end{equation}
where $N$ denotes the total number of cells adjacent to the cell $({x_0},{y_0})$. Then, the normal and tangential derivatives of the macroscopic variables can be computed from the known derivative terms $\bm{b}_x$ and $\bm{b}_y$. Finally, the concerned derivative terms (e.g. $a_n^l$, $a_n^r$) can be computed by following the detailed process reported in Ref.~\cite{xu2001gas}.

For compressible flow simulations, the limiter is often introduced to compute the gradients to avoid a huge value and/or to eliminate the possible spurious oscillations. In this paper, the minmod limiter~\cite{sweby1984high,de1993quadtree} is used for updating the derivatives of macroscopic variables,
\begin{equation}\label{Eq09}
\bm{\phi}_W  = \min \left( 1, {\frac{{\left| {\bm{W}^c - {{\max }_{path}}\left( \bm{W} \right)} \right|}}{{\left| {\bm{W}^c - {{\max }_{node}}\left( {\bm{W}} \right)} \right|}}}, {\frac{{\left| {\bm{W}^c - {{\min }_{path}}\left( {{\bm{W}}} \right)} \right|}}{{\left| {\bm{W}^c - {{\min }_{node}}\left( {{\bm{W}}} \right)} \right|}}}
\right),
\end{equation}
where the ${{\min }_{path}\left( \bm{W} \right)}$ and ${{\max }_{path}}\left( \bm{W} \right)$ are the minimum and maximum value of $\bm{W}$ around the cell, which are used to calculate the gradients of $\bm{W}$ at the cell. The ${{\min }_{node}}\left( \bm{W} \right)$ and ${{\max }_{node}}\left( \bm{W} \right)$ are the minimum and maximum value of $\bm{W}$ at the node of the cell, which are reconstructed by the gradients. $\bm{W}^c$ represents the macroscopic variables at the center of the cell. Note that, all the macroscopic variables $\bm{W}$ have their own minmod variety coefficients $\bm{\phi}_W$, and Eq.~(\ref{Eq06}) can be rewritten as
\begin{equation}\label{Eq10}
\bm{W} = \bm{W}_0 + \bm{\phi}_W \left( {{\bm{b}_x}\left( {x - {x_0}} \right) + {\bm{b}_y}\left( {y - {y_0}} \right)} \right) .
\end{equation}

To determine the time derivative terms, ${A^l}$, ${A^r}$ and $\bar A$, the flow conservation constraints must be utilized: the non-equilibrium part of the gas distribution function should have no direct contribution to the macroscopic conservative variable
\begin{equation}\label{Eq11}
\begin{array}{l}
\int {\left( {a_n^lu + a_t^lv + {A^l}} \right) \bm{\psi} {g^l}d\Xi }  = \bm{0},\\
\int {\left( {a_n^ru + a_t^rv + {A^r}} \right) \bm{\psi} {g^r}d\Xi }  = \bm{0},
\end{array}
\end{equation}
and the macroscopic variables mass, momentum, and energy should be conserved during particle collisions,
\begin{equation}\label{Eq12}
\int _0^{\Delta t} \int {\bm{\psi} \left( {f - g} \right)d\Xi dt} = \bm{0}.
\end{equation}
Thus, the systems of linear equations about those derivative terms can be built, and can be solved by the Gaussian elimination.

Now, the gas distribution function at the cell interface can be expressed as
\begin{equation}\label{Eq13}
\begin{aligned}
{f_{cf}} = &\left( {1 - {e^{ - \frac{t}{\tau }}}} \right){g_0}\\
& + \left( {\tau \left( { - 1 + {e^{ - \frac{t}{\tau }}}} \right) + t{e^{ - \frac{t}{\tau }}}} \right)\left( {\left( {\bar a_n^lH[u] + \bar a_n^r\left( {1 - H[u]} \right)} \right)u + {{\bar a}_t}v} \right){g_0}\\
& + \tau \left( {\frac{t}{\tau } - 1 + {e^{ - \frac{t}{\tau }}}} \right)\bar A{g_0}\\
& + {e^{ - \frac{t}{\tau }}}\left( {1 - \left( {t + \tau } \right)\left( {ua_n^l + va_t^l} \right)} \right)H[u]{g^l}\\
& + {e^{ - \frac{t}{\tau }}}\left( {1 - \left( {t + \tau } \right)\left( {ua_n^r + va_t^r} \right)} \right)\left( {1 - H[u]} \right){g^r}\\
& + {e^{ - \frac{t}{\tau }}}\left( { - \tau {A^l}H[u]{g^l} - \tau {A^r}\left( {1 - H[u]} \right){g^r}} \right) ,
\end{aligned}
\end{equation}
where the Heaviside function reads
\begin{equation}\nonumber
H[x] = \left\{ \begin{array}{l}
0 \qquad  x < 0,\\1 \qquad  x \ge 0. \end{array} \right.
\end{equation}

Finally, we can obtain the flux term $\bm{F}$ of the macroscopic variables across the cell interface from the gas distribution function at the cell interface, e.g., the flux in x-direction can be computed as
\begin{equation}\label{Eq14}
{\bm{F}} = \int_0^{\Delta t} {\int {\bm{\psi} u{f_{cf}}d\Xi } dt} .
\end{equation}

\subsection{MATRIX-FREE LU-SGS SCHEME}\label{sec:LUSGS}

Generally, the explicit time marching of finite volume method for the conserved variables are written as
\begin{equation}\label{Eq15}
{\bm{W}^{n + 1}} = {\bm{W}^n} - \frac{1}{\Omega }\sum\limits_{k = 1}^N {\bm{F}_k \cdot \bm{S}_k} ,
\end{equation}
where $\Omega$ is the volume of the control volume. $\bm{S}$ represent the area of the interfaces of this control volume.

In this study, considering an implicit scheme for the time marching. Firstly, Eq.~(\ref{Eq15}) can be rewritten as
\begin{equation}\label{Eq16}
\frac{{{\bm{W}^{n + 1}} - {\bm{W}^n}}}{{\Delta t}}\Omega  = - \frac{1}{{\Delta t}}\sum\limits_{k = 1}^N {\bm{F}_k \cdot \bm{S}_k} .
\end{equation}
Following the implicit algorithm presented in Ref.~\cite{li2014implicit}, for simplicity, the right hand term of Eq.~(\ref{Eq16}) is referred to as a residual term $\bm{M} = - \frac{1}{{\Delta t}}\sum\limits_{k = 1}^N {\bm{F}_k \cdot \bm{S}_k}$, and Eq.~(\ref{Eq16}) can be expressed as the differential form
\begin{equation}\label{Eq17}
\frac{{\partial {\bm{W}_i}}}{{\partial t}}{\Omega _i} = {\bm{M}_i} .
\end{equation}
Then, above equation can be further rewritten as
\begin{equation}\label{Eq18}
\frac{{\partial {\bm{W}_i}}}{{\partial t}}{\Omega _i} = {\left( {\frac{{\partial \bm{M}}}{{\partial \bm{W}}}} \right)_i^n}\Delta \bm{W_i} + {\bm{M_i}^n} ,
\end{equation}
where the term $\frac{{\partial \bm{M}}}{{\partial \bm{W}}}$ is a Jacobian matrix, which takes the form in first-order Roe's scheme in which the Roe's flux can be linearized~\cite{chen2000fast}. According to the idea of LU-SGS method, we first split this Jacobian matrix into three parts: lower triangular matrix, upper triangular matrix and diagonal terms. Thus, Eq.~(\ref{Eq18}) can be rewritten as
\begin{equation}\label{Eq19}
\left( {\bm{L} + \bm{U} + \bm{D}} \right)\Delta \bm{W} = {\bm{M}^n} ,
\end{equation}
with
\begin{equation}\label{Eq20}
\left\{ \begin{array}{l}
\bm{L} = \sum\limits_{j < i } {\left( {\frac{1}{2}\frac{{\partial \bm{F}\left( {\bm{W}} \right)}}{{\partial \bm{W}}} \cdot \bm{S} - {{\left( {{\Lambda _c}} \right)}_{ij}}} \right)} ,\\
\bm{U} = \sum\limits_{j > i } {\left( {\frac{1}{2}\frac{{\partial F\left( {\bm{W}} \right)}}{{\partial \bm{W}}} \cdot \bm{S} - {{\left( {{\Lambda _c}} \right)}_{ij}}} \right)} ,\\
\bm{D} = \left( {\frac{{{\Omega _i}}}{{\left( {\Delta t} \right)_i^{imp}}} + \sum\limits_{j \in N\left( i \right)} {{{\left( {{\Lambda _c}} \right)}_{ij}}} } \right)\bm{I} ,
\end{array} \right.
\end{equation}
where ${\Lambda _c}$ reads
\begin{equation}\label{Eq21}
\left( \Lambda _c \right)_i = \sum\limits_{j = 1}^N {\left( \left| {\vec V}_i \cdot {\vec n}_j \right| + {C_i} \right){S_j}} ,
\end{equation}
with ${\vec V}_i$ represents velocity vector, $C_i$ denotes the sound speed at the cell $i$, and $S_j$ is area of the interface $j$.

Then, the LU-SGS scheme is applied to solve Eq.~(\ref{Eq19}):
\begin{equation}\label{Eq22}
\left( {\bm{L} + \bm{D}} \right){\bm{D}^{ - 1}}\left( {\bm{D} + \bm{U}} \right)\Delta \bm{W} = {\bm{M}^n} .
\end{equation}

When computing the RHS term at time n-level, the n-level's explicit time step should be known. Here, this explicit time step is computed by
\begin{equation}\label{Eq23}
\Delta t = \min \left( {\delta_{CFL_0}\frac{{{\Omega _i}}}{{{{\left( {{\Lambda _c}} \right)}_i}}}} \right) .
\end{equation}
To accelerate convergency (time marching), an implicit time step is introduced with the similar form,
\begin{equation}\label{Eq24}
\left( {\Delta t} \right)_i^{imp} = {\delta_{CFL_{imp}}}\frac{{{\Omega _i}}}{{{{\left( {{\Lambda _c}} \right)}_i}}} .
\end{equation}

The implicit CFL number $\delta_{CFL_{imp}}$ can be adjusted with the decreasing residual. We choose infinity norm of residual to be the criterion that how to adjust the implicit CFL number~\cite{tidriri1997preconditioning}.
\begin{equation}\label{Eq25}
{\delta_{CFL_{imp}}^{n + 1}} = {\delta_{CFL_{imp}}^{n}} \frac{{{{\left\| {{\bm{M}^n}} \right\|}_\infty }}}{{{{\left\| {{\bm{M}^{n + 1}}} \right\|}_\infty }}} .
\end{equation}

\subsection{TURBULENCE MODEL}\label{sec:TURBULENCE MODEL}
In order to solve the turbulent flow problems, the SA one-equation turbulence model is employed, which solves a transport equation for a transformed eddy viscosity $\tilde \nu$. In this study, the governing transport equation followed in the work by Moro et al.~\cite{moro2011navier} reads
\begin{equation}\label{Eq26}
\frac{{D\tilde \nu }}{{Dt}} = {C_{b1}}\tilde S\tilde \nu  - {C_{w1}}{f_w}{\left( {\frac{{\tilde \nu }}{d}} \right)^2} + \frac{1}{\sigma }\left[ {\left( {\nabla  \cdot \left( {\left( {\nu  + \tilde \nu } \right)\nabla \tilde \nu } \right)} \right) + {C_{b2}}{{\left( {\nabla \tilde \nu } \right)}^2}} \right],
\end{equation}
where the three terms at the right hand side represent turbulence production, turbulence destruction due to wall and turbulence diffusion/propagation, respectively. For the computation/definitions of the eddy viscosity $\nu_t$ and all other parameters and constants involved in the SA model, please find the details in Ref.~\cite{moro2011navier}. Here, we focus on introducing how to solve this SA transport equation on unstructured hybrid mesh by FVM.

Following the approach proposed in Ref.~\cite{shenderobust}, a semi-discrete form in cell $i$ for the discretization of SA model equation reads
\begin{equation}\label{Eq27}
\frac{{\partial {{\tilde \nu }_i}}}{{\partial t}}{\Omega _{i}} = {\Omega _{i}}{Sr_i} + {R_i}  ,
\end{equation}
where $\Omega _i$ is volume of element $i$, $Sr_i$ denotes the source terms,
\begin{equation}\label{Eq28a}
S{r_i} = {C_{b1}}\tilde S\tilde \nu  - {C_{w1}}{f_w}{\left( {\frac{{\tilde \nu }}{d}} \right)^2},
\end{equation}
and $R_i$ denotes the flux term which including three parts: inviscid flux $R_{inv}$, viscous flux $R_{vis}$ and anti-diffusion flux $R_{ad}$,
\begin{equation}\label{Eq28}
{R_{inv}} =  - \sum\limits_{k = 1}^N {{F_{k,inv}}{S_k}} ,~
{R_{vis}} = \sum\limits_{k = 1}^N {{F_{k,vis}}{S_k}} ,~
{R_{ad}} =  - \sum\limits_{k = 1}^N {{F_{k,ad}}{S_k}} ,
\end{equation}
where $S_k$ represents the area of the $k$th interface of cell $i$ and we define
\begin{equation}\label{Eq29}
\begin{array}{l}
{F_{k,inv}} = u_{ \bot i}^ + {{\tilde \nu }_i} + u_{ \bot j}^ - {{\tilde \nu }_j},\\
{F_{k,vis}} = \frac{1}{{2\sigma }}\left[ {\left( {{\nu _i} + {\nu _j}} \right) + \left( {1 + {C_{b2}}} \right)\left( {{{\tilde \nu }_i} + {{\tilde \nu }_j}} \right)} \right]{\left( {\nabla \tilde \nu  \cdot \vec n} \right)_k},\\
{F_{k,ad}} = \frac{{{C_{b2}}}}{\sigma }{{\tilde \nu }_i}{\left( {\nabla \tilde \nu  \cdot \vec n } \right)_k},
\end{array}
\end{equation}
with
\begin{equation}\label{Eq30}
u_ \bot ^ \pm  = \frac{1}{2}\left( {{\vec V} \cdot \vec n \pm \left| {{\vec V} \cdot \vec n} \right|} \right) ,~
{\left( {\nabla \tilde \nu  \cdot \mathord{\buildrel{\lower3pt\hbox{$\scriptscriptstyle\rightharpoonup$}}
\over n} } \right)_k} = \frac{{\left( {{{\tilde \nu }_j} - {{\tilde \nu }_i}} \right)}}{{\left| {{{\vec r}_{ij}} \cdot {\vec n} } \right|}} .
\end{equation}
Note that, in above two equations, ${\vec r_{ij}}$ represents the distance vector from the center of cell $i$ point to the center of cell $j$ which shares the $k$th interface and $\vec n$ denotes unit normal vector of the cell interface.

Now, Eq.~(\ref{Eq27}) can be rewritten as the following implicit form
\begin{equation}\label{Eq31}
\left( {\frac{{{\Omega _{i}}}}{{\Delta t}} - \frac{{\partial R}}{{\partial \tilde \nu }} - {\Omega _{i}}\frac{{\partial Sr}}{{\partial \tilde \nu }}} \right)\Delta \tilde \nu  = {\Omega _{i}}{Sr}^n + {R^n}.
\end{equation}
For the derivative terms, taking cell $i$ and its neighbors $j$ into consideration, Eq.~(\ref{Eq31}) becomes
\begin{equation}\label{Eq32}
\left( {\frac{{{\Omega _{i}}}}{{\Delta t}} - \frac{{\partial {R_i}}}{{\partial {{\tilde \nu }_i}}} - {\Omega _{i}}\frac{{\partial {Sr_i}}}{{\partial {{\tilde \nu }_i}}}} \right)\Delta {\tilde \nu _i} - \frac{{\partial {R_i}}}{{\partial {{\tilde \nu }_j}}}\Delta {\tilde \nu _j} =  {\Omega _{i}}{Sr}_i^n + {R_i^n}.
\end{equation}
Here, we can obtain a system of implicit linear equations which can be solved by using the LU-SGS scheme.

In BGK model~\cite{bhatnagar1954model}, the particle average collision time equal to the ratio of the dynamic viscosity to the static pressure,
\begin{equation}\label{Eq33}
\tau  = \frac{\mu }{p} .
\end{equation}
In order to couple the turbulence effect into the present GKS model, the computation of relaxation time should be modified by incorporating the eddy viscosity into the total viscosity~\cite{xiong2011numerical}
\begin{equation}\label{Eq34}
\tau  = \frac{{{\mu _l} + {\mu _t}}}{p} ,
\end{equation}
where ${\mu _l}$ and $\mu _t$ denote the laminar dynamic viscosity and the eddy dynamic viscosity, respectively, and $\mu _t$ is computed from the updated values of the density and the eddy viscosity $\nu_t$, $\mu _t = \rho \nu_t$.

\subsection{NON-MANIFOLD HYBRID MESH DATA STRUCTURE}\label{sec:NMHM}
For the numerical simulation on unstructured hybrid mesh, an efficient mesh data structure is required to access all mesh data without causing high cost. In the present study, the non-manifold hybrid mesh data (NHMD)~\cite{ebeida2009mesh} is applied, in which six kinds of elements are considered: nodes, lines, edges, faces, cell and entity. These elements form a library and can be indexed each other~\cite{ebeida2009mesh,FLD4239}. Thus, we do not need to repeat the process of searching for geometrical information.

For example, as shown in Fig.~\ref{fig:Fig02}, NHMD is categorized into three parts: 1) the number of nodes and every kind of cells; 2) the locations of all nodes denoted by $X$, $Y$, $Z$; 3) the connectivity of every cell represented by the index number of all vertex of this cell. By using this method, we can easily convert arbitrary hybrid mesh into this kind of mesh data for the present IGKS solver since there is no requirement for the geometric features of every cell.

\subsection{BOUNDARY CONDITIONS}\label{sec:BOUNDARY CONDITION}
Following the ``ghost cell" method shown in Xu's studies~\cite{xu2001gas}, this subsection is mainly about how to implement this idea numerically for the present IGKS. Here, four kinds of boundary conditions including free-stream, symmetric, non-slip wall and outflow boundary conditions are introduced, and both physical values and their increments are considered due to the used implicit scheme.

\subsubsection{Free-stream boundary condition}
For the physical variables including four conservative variables $W_i$ and the transformed eddy viscosity $\tilde \nu$ , simply set as
\begin{equation}\nonumber
\begin{aligned}
{\left( \rho  \right)_{ - 1}} = {\rho _0} ,~{\left( {\rho U} \right)_{ - 1}} = {\rho _0}{U_0} ,~{\left( {\rho V} \right)_{ - 1}} = {\rho _0}{V_0} ,~{\left( \rho E \right)_{ - 1}} = {\rho E_0} ,~{\left( {\tilde \nu } \right)_{ - 1}} = {{\tilde \nu }_0} .
\end{aligned}
\end{equation}

For increments of these physical variables,
\begin{equation}\nonumber
\begin{aligned}
{\left( {\Delta \rho } \right)_{ - 1}} = {\left( {\Delta \rho U} \right)_{ - 1}} = {\left( {\Delta \rho V} \right)_{ - 1}} = {\left( {\Delta \rho E} \right)_{ - 1}} = {\left( {\Delta \tilde \nu } \right)_{ - 1}} = 0 .
\end{aligned}
\end{equation}

\subsubsection{Symmetric boundary condition}

For the physical variables,
\begin{equation}\nonumber
\begin{aligned}
{\left( \rho  \right)_{ - 1}} = {\left( \rho  \right)_1} ,~{\left( {\rho {V_n}} \right)_{ - 1}} =  - {\left( {\rho {V_n}} \right)_1} ,~{\left( {\rho {V_t}} \right)_{ - 1}} = {\left( {\rho {V_t}} \right)_1} ,~{\left( \rho E \right)_{ - 1}} = {\left( \rho E \right)_1} ,~{\left( {\tilde \nu } \right)_{ - 1}} = {\left( {\tilde \nu } \right)_1} .
\end{aligned}
\end{equation}

Similarly, for increments of the physical variables,
\begin{equation}\nonumber
\begin{aligned}
{\left( {\Delta \rho } \right)_{ - 1}} &= {\left( {\Delta \rho } \right)_1} ,~{\left( {\Delta \rho {V_n}} \right)_{ - 1}} =  - {\left( {\Delta \rho {V_n}} \right)_1} ,~{\left( {\Delta \rho {V_t}} \right)_{ - 1}} = {\left( {\Delta \rho {V_t}} \right)_1} ,\\
{\left( {\Delta \rho E} \right)_{ - 1}} &= {\left( {\Delta \rho E} \right)_1} ,~{\left( {\Delta \tilde \nu } \right)_{ - 1}} = {\left( {\Delta \tilde \nu } \right)_1} .
\end{aligned}
\end{equation}

\subsubsection{Non-slip boundary condition}

For the physical variables,
\begin{equation}\nonumber
\begin{aligned}
{\left( \rho  \right)_{ - 1}} = {\left( \rho  \right)_1} ,~{\left( {\rho U} \right)_{ - 1}} =  - {\left( {\rho U} \right)_1} ,~{\left( {\rho V} \right)_{ - 1}} =  - {\left( {\rho V} \right)_1} ,~{\left( \rho E \right)_{ - 1}} = {\left( \rho E \right)_1} ,~{\left( {\tilde \nu } \right)_{ - 1}} =  - {\left( {\tilde \nu } \right)_1} .
\end{aligned}
\end{equation}

For increments of the physical variables,
\begin{equation}\nonumber
\begin{aligned}
{\left( {\Delta \rho } \right)_{ - 1}} &= {\left( {\Delta \rho } \right)_1} ,~{\left( {\Delta \rho U} \right)_{ - 1}} =  - {\left( {\Delta \rho U} \right)_1} ,~{\left( {\Delta \rho V} \right)_{ - 1}} =  - {\left( {\Delta \rho V} \right)_1} ,\\
{\left( {\Delta \rho E} \right)_{ - 1}} &= {\left( {\Delta \rho E} \right)_1} ,~{\left( {\Delta \tilde \nu } \right)_{ - 1}} =  - {\left( {\Delta \tilde \nu } \right)_1} .
\end{aligned}
\end{equation}

\subsubsection{Outflow boundary condition}
For the outflow boundary conditions, the Riemann invariants presented by Carlson~\cite{carlson2011inflow} are introduced in the present treatment of outflow boundary condition to reduce the disturbance propagating into flow field. As reported in Ref.~\cite{carlson2011inflow}, if the physical state (outside the domain) is on the right side of boundary and the solution space (inside the domain) is on its left, we define two Riemann invariants correspond to the incoming $R^-$ and outgoing $R^+$ characteristic waves, respectively
\begin{equation}\nonumber
\begin{aligned}
{R^ - } = {U_o} - \frac{{2{c_o}}}{{\gamma  - 1}} ,\\
{R^ + } = {U_i} + \frac{{2{c_i}}}{{\gamma  - 1}} ,
\end{aligned}
\end{equation}
where the two subscripts $o$ and $i$ represents the normal values/components outside and inside the flow field, respectively. For the subsonic outflow boundary conditions, both incoming and outgoing characteristic waves exist, so the velocity $U_b$ and the speed of sound $c_b$ at the boundary are computed as the sum and difference of the invariants, respectively
\begin{equation}\nonumber
{U_b} = \frac{1}{2}\left( {{U_R^ + } + {U_R^ - }} \right) ,~{c_b} = 4\left( {\gamma  - 1} \right)\left( {{U_R^ + } - {U_R^ - }} \right) .
\end{equation}
Then, the density and energy on the boundary can be calculated as follows
\begin{equation}\nonumber
\rho_b = \rho_i, {\left( \rho E \right)}_b = \frac{1}{2}{\rho _b}\left( {U_b^2 + V_i^2 + \frac{{K + 2}}{{2{\lambda _i}}}} \right),
\end{equation}
where $i$  represents inside the flow field, and the transformed eddy viscosity on the boundary just set as $\tilde \nu_b = \tilde \nu_i$.

For the increments of macroscopic variables, all are simply set at zero. This strategy does not produce negative effect in convergence.

\section{NUMERICAL RESULTS}\label{sec:NUMERICAL RESULTS}
In this section, we perform some numerical simulations to validate the present IGKS under unstructured hybrid mesh in different types of fluid flow; from laminar to turbulent flows. In all simulations of steady flows, the convergent criterion towards the steady state is set as follows
\begin{equation}\label{Eq40}
\max \limits_{k \in Nc} \left( {\frac{ \left| {{\bf{W}}_k^{n+1}}-{{\bf{W}}_k^n}\right| }{\left| {{\bf{W}}_k^n} \right|} }\right) \le 3 \times {10^{ - 8}},
\end{equation}
where $Nc$ is the number of mesh cells in fluid field.

\subsection{Lid-driven cavity flow on uniform and unstructured hybrid meshes}
The 2D lid-driven cavity flow has been studied by many researchers and employed as a benchmark case of incompressible viscous flow. In this part, the numerical simulations for the 2D lid-driven cavity flow on uniform and unstructured hybrid meshes are performed.

The flow configuration of this case is consists of a 2D square cavity with its top plate moving with a uniform horizontal velocity $\bm{U_0}$, while other three walls are set to be static non-slip boundary. In this series of simulations, the Reynolds number is defined as $Re = \rho U_0 L_{ref}/\mu$ and set at $1,000$ unless otherwise notified, in which the height/width of the cavity is defined as the reference length, ${L_{ref}}=1.0$, and the Mach number is set to be $0.1\sqrt 3 $. Besides, the $100 \times 100$ uniform grid and an unstructured hybrid grid with a total number of $8933$ mesh cells, are respectively used for the present two simulations. Here, the unstructured hybrid grid is composed of three kinds of cells, quadrilaterals, pentagons, and hexagons, as shown in Fig.~\ref{fig:Fig03}. In the present 2D lid-driven cavity flow simulation with the unstructured hybrid grid, the mesh shown in Fig.~\ref{fig:Fig03} is the final adaptive mesh which are generated and controlled by performing refinement in the region where the gradients and vorticities $\omega$ are larger than their mean level.

Fig.~\ref{fig:Fig04} shows the two computed streamlines (stream functions $\psi$). It is clear that the same flow structures are obtained from the two simulations with different meshes, which are in good agreement with those results shown in previous literatures~\cite{ghia1982high,zhuo2012filter,hou1995simulation}. Fig.~\ref{fig:Fig05} presents the comparisons of the computed velocity profiles along the mid-height ($y/L_{ref}=0.5$) of the cavity against the numerical results of multi-grid method reported by Ghia et al.~\cite{ghia1982high}. It can be found from this plot that both results of the present simulations with uniform and unstructured hybrid grids agree well with the benchmark data of Ghia et al.~\cite{ghia1982high}.

In order to demonstrate the accuracy of the present IGKS, the locations $(x,y)$ of the primary and secondary vortices, as well as the values of vorticity at these locations are grouped in Table~\ref{table:Table1}. The values of locations and vorticity are normalized by $L_{ref}$ and $U_0/L_{ref}$, respectively. For purpose of comparison, the numerical results reported by Ghia et al.~\cite{ghia1982high} and two lattice Boltzmann method (LBM) results obtained with uniform grid of $255 \times 255$~\cite{zhuo2012filter,hou1995simulation} are also grouped into this table. The present comparisons show that, both IGKS results obtained with uniform and unstructured hybrid grids are in good agreement with those results given in the previous literatures~\cite{ghia1982high,zhuo2012filter,hou1995simulation}: the relative differences between the present results and those benchmark data of Ghia et al.~\cite{ghia1982high} are less than $10.4\%$, $2.0\%$ and $2.8\%$ for the values of locations, vorticity $\omega$ and stream function $\psi$, respectively.

To further compare the present results obtained with two different kinds of meshes, Fig.~\ref{fig:Fig06} shows the computed pressure contours of two simulations at the same contour levels. It is clear that, the two pressure contour lines agree well in the most flow regions, even near the top-right corner in which there is singular point on the top-moving wall. Also, there are a slight differences between both pressure contour lines at the primary vortex, near the wall and the top-left corner. Overall, this comparison shows that the present IGKS with unstructured hybrid grids can give us the satisfactory results.

For the present two simulations with different kinds of meshes, the variations of the residual value determined by Eq.~(\ref{Eq40}) and the CFL number over the iteration step are grouped in Fig.~\ref{fig:Fig07} and Fig.~\ref{fig:Fig08}, respectively. It can be observed from the two plots that, the two simulations with uniform and unstructured hybrid grids reach the steady state quickly and only take about $4,430th$ and $4,270th$ iteration steps, respectively,, and both variable CFL numbers grow very fast with setting the initial implicit CFL number $\delta_{CFL_0} = 1,000$ and their final values increase up to the order of one million.

When compared to the explicit GKS, it is no doubt that implicit iteration in present IGKS needs more computational time (takes $100\%$ more time than the explicit GKS per iteration step, which can be calculated from the data grouped in Table~\ref{table:Table2}) to complete one iteration step, but IGKS can achieve convergence much faster in the simulations of steady flows since the quite large CFL number can be used in the IGKS simulations. To show the advantages of the present IGKS and assess its computational efficiency, another simulation is also performed by using the standard explicit GKS with the uniform mesh for the purpose of comparisons. Fig.~\ref{fig:Fig09} shows the comparison of convergency curves for the present IGKS and explicit GKS simulations. It is clear that, the IGKS simulation only take $4,430$ iteration steps to converge while about $608,000$ iteration steps are need for the explicit GKS simulation. This result is consistent with the work of Zhu et al (2016) which is based on the unified gas kinetic scheme~\citep{zhu2016implicit}. Considering the computational efficiency, the computational time of the present IGKS and the explicit GKS simulations on uniform and unstructured hybrid grid have been computed and grouped in Table~\ref{table:Table2}. It is noted that, the total numbers of mesh cells and the cost of computational time for running the corresponding iteration steps are included here for all simulation cases performed in this paper. Table~\ref{table:Table2} shows that, for the present steady cavity flow cases, the IGKS simulations are about two orders of magnitude faster than the explicit GKS simulations. Meanwhile, due to few mesh cells used for the unstructured hybrid grid, the computational time of the present simulations of cavity flows on unstructured hybrid grid is just about $97.2\%$ of those taken by the corresponding simulations on uniform grid.

In all, it can be concluded from these comparisons that, the present IGKS with unstructured hybrid grid not only has capability for accurately predicting the steady incompressible viscous flows, but also has good computational efficiency compared with the explicit GKS.

\subsection{Laminar flows around a zero-pressure-gradient flat plate}
In this part, we consider a laminar boundary flow with Mach number $Ma=0.2$ and $Re=5,000$ over a flat plate. In Fig.~\ref{fig:Fig10} the schematic view of the computational domain is given along with the corresponding physical boundary conditions. It is noted that, the length of flat plate is set to be the characteristic length, $L_{ref}=1.0$, and the leading edge point locates at $(0.0, 0.0)$; the inflow at the left boundary is set as uniform with a constant horizontal velocity, and the Riemann invariants are considered for the upper boundary and the outflow boundary; for bottom boundary, the symmetric and non-slip boundary conditions are considered.

As we known, the mesh resolution for boundary layer is crucial, thus using hybrid mesh is a wise strategy to enhance mesh efficiency and reduce computational cost as low as possible. As shown in Fig.~\ref{fig:Fig11a}, a hybrid mesh with $13441$ cells is used in this simulation, and the local view near wall is plotted in Fig.~\ref{fig:Fig11b}. In Fig.~\ref{fig:Fig12}, the computed U and V velocity profiles at three locations $x=0.202,~x=0.503$ and $x=0.700$ are plotted, where $\eta  = y\sqrt {\frac{{{U_\infty }}}{{\nu x}}} $ is the dimensionless distance from the flat plate and the exact Blasius solutions are also grouped into these plots. Fig.~\ref{fig:Fig13} shows the computed skin friction coefficient compared with the exact Blasius solution. In those comparisons, the present IGKS results of the velocity profiles and the skin friction coefficient profile fit the exact solutions very well.

Besides, for this case, the IGKS simulation converges at $4,320$th iteration steps and the corresponding computational time is about $717.4$ seconds, and which also shows about two order of magnitude improvement in the computational times when compared to that of explicit GKS, as shown in Table~\ref{table:Table2}. The history curves of the CFL number and residual value are plotted in Fig.~\ref{fig:Fig14}.

\subsection{Turbulent flow around a zero-pressure-gradient flat plate}
In this part, to validate the present IGKS-SA method in simulation of turbulent flows, we continue to investigate the turbulent boundary layer case with Mach number $Ma=0.2$ and $Re=5 \times 10^6$ over a flat plate. The delicate flow structures in boundary layer, such as velocity profile, skin friction coefficient and eddy viscosity, are computed for comparisons.

In this test, the length of the flat plate is 2.0, the height of the whole flow field is 1.0, and the
leading edge of the flat plate is at $x=0$. The final refined mesh reported in our previous simulations with the explicit GKS~\cite{FLD4239} is used in the present simulation, where the total number of cells is $26,386$ and most of the cells gather within the boundary layer, as shown in Fig.~\ref{fig:Fig15a}. Near the wall boundary, the high resolution of mesh is guaranteed where the first grid point off the wall is located at $y^+\approx1.1$, the local view of mesh near the wall is shown in Fig.~\ref{fig:Fig15b}.

For this case, all boundary conditions used in above laminar boundary layer case are still applied. For the SA turbulence model, we set $\tilde \nu  = 0.1\nu $ for the left inflow boundary and the initial flow field~\cite{allmaras2012modifications}, while set $\tilde \nu = 0$ for the top, outflow and wall boundaries.

Fig.~\ref{fig:Fig16a} and~\ref{fig:Fig16b} show the present computed velocity profiles at two positions $x = 0.97008$ and $x = 1.90334$, respectively. The benchmark solutions of CFL3D reported by Wilcox~\cite{LangleyResearchCenter} are grouped into the two plots for the purpose of comparisons. It is clear that, both profiles computed by the present IGKS are in good agreement with the benchmark data of CFL3D. Note that, in Wilcox's study~\cite{LangleyResearchCenter}, the number of cells used in the simulation is $545 \times 385$ which is about eight times the present computational grid. In Fig.~\ref{fig:Fig17} and Fig.~\ref{fig:Fig18}, we illustrate the present results of the skin friction coefficient and the dimensionless eddy viscosity profile at $x=0.97$, respectively, which also agree well with those benchmark solutions of CFL3D~\cite{LangleyResearchCenter}.

In Fig.~\ref{fig:Fig19}, we plot the variations of CFL number and residual value over the iteration. The present IGKS simulation can reach convergency at about $5,340th$ iteration steps, while for the previous explicit GKS simulation, it takes up to about $810,000$ iteration steps (see Table~\ref{table:Table2}). To investigate the computational efficiency of the present IGKS for the turbulent flows, the computational time of the present IGKS simulation and the previous simulation of the explicit GKS are also included into Table~\ref{table:Table2}, which shows that the present IGKS simulation only takes $1.4\%$ of the computational time of the explicit GKS simulation. It should be pointed out that, compared to the value of $90.2$ observed from the laminar flow cases, the ratio of computational time ($72.3$) between explicit GKS and IGKS in the turbulent flow simulations is a little bit smaller since the implicit scheme is also considered for the SA turbulence model.

\subsection{Turbulent flow around NHLP multi-element airfoil}
In order to validate the applicability of the present IGKS-SA solver for predicting the complex turbulent flow with complex boundaries, a turbulent flow around a two-dimensional supercritical airfoil with high-lift devices, NHLP-2D three-element airfoil, is investigated in this study.

The configuration of NHLP-2D includes a $12.5\%C$ leading-edge slat and a $33\%C$ single slotted-flap, where $C$ is the chord length of the nested configuration. Fig.~\ref{fig:Fig20} shows this geometry with the computational grid near airfoil. It is clear that, to reduce the total number of mesh cells, the cell scale increased fast within the region out of boundary layer and the outer of wake region. However, in boundary layer and near-wall wake regions, the fine grids are used in this study; as shown in Figs.~\ref{fig:Fig21a} and~\ref{fig:Fig21b}, the body-fitted quadrilaterals are applied to capture the thin boundary layer and the fine triangles are used to fill the two gaps between slat and flap with the main element. Here, the first node point off the airfoil surface for the finest grid is located such that $y^+ \scriptsize{\sim} 0.3$, and the total number of mesh cells used in this simulation is $64,344$. The flow conditions are Mach number $Ma=0.195$, Reynolds number $Re=3.52 \times 10^6$ and an angle of attack of $4.01^\circ$.

The contours of density, pressure, Mach number and eddy viscosity computed from the present IGKS-SA solver are shown in Figs.~\ref{fig:Fig22},~\ref{fig:Fig23},~\ref{fig:Fig24} and~\ref{fig:Fig25}, respectively. It can be found from those plots that, the slat wake merges with the boundary layer of main element and seems to be disappearing near the trailing edge of main element. Fig.~\ref{fig:Fig26} presents the surface pressure coefficients calculated with present IGKS-SA solver, which are in good agreement with the experiment data reported by Morrison~\cite{morrison1998numerical}. It is important to point out that the present SA turbulence model is not calibrated to predict transition, and actually, the fully turbulent simulation is performed here.

In Fig.~\ref{fig:Fig27}, the variations of CFL number and residual value over the iteration are illustrated. The present IGKS simulation can reach convergence at $3,650th$ iteration steps, while for the previous explicit GKS simulation, it needs about {\color{red}{$520,000$}} iteration steps (see Table II). To investigate the computational efficiency of the present IGKS for the turbulent flows, the computational time of the present IGKS simulation and the previous simulation of the explicit GKS are also included into Table II; the present IGKS also exhibits $94.5$ times improvement than the classic explicit GKS.

In all, the present IGKS solver is capable of the laminar flow simulations on both regular structured grid and unstructured hybrid mesh, and when coupled with the turbulence model, e.g. SA model, the present IGKS solver shows good convergence behavior and accuracy for simulating the complex turbulent flow with complex boundaries which contains arbitrary complex configurations.

\section{Conclusions}\label{sec:conlustion}
This paper presents an implicit GKS coupled with Splalart-Allmaras turbulence model based on LU-SGS scheme under unstructured hybrid mesh. As a kind of efficient mesh data structure for unstructured hybrid mesh, an extended NHMD is used in this study to reduce the extra computational costs. In this paper, to validate the capability of the present IGKS solver and show the advantage of the present IGKS compared with the explicit GKS, four steady viscous flow cases including lid-driven cavity flows, laminar and turbulent boundary flows over a flat plate, and the turbulent flow around NHLP-2D airfoil are performed.

For the lid-driven cavity flows, both simulations on uniform and unstructured hybrid mesh are considered, and almost the same results have been obtained by the present IGKS solver which also agree well with the benchmark data reported by other researchers. In the present simulation for the laminar boundary flow over a flat plate, the velocity profiles at three different locations and the skin friction coefficient profiles along the flat plate computed from the present IGKS solver agree quite well with the Blasius analytical solutions; also, for the turbulent boundary flow case, the delicate flow structures in boundary layer are predicted by the present IGKS coupled with SA model, show us the satisfactory results compared with the available data of CFL3D. At last, the steady turbulent flow around the NHLP-2D airfoil is considered to validate the capability of the present IGKS-SA solver for predicting the complex flow with complex configurations. The merging of the slat wake with the main element boundary layer are clearly observed from the present simulation and the present results of surface pressure coefficients are also in good agreement with the available experimental data. Besides, the comparisons of computational time for the steady flow simulations, between the present IGKS and explicit GKS on uniform and/or unstructured hybrid mesh, have been carried out, indicating that the present IGKS with unstructured hybrid mesh have the best computational efficiency for steady state solutions to get the convergent flow field.

Through this study, the IGKS coupled with SA turbulence model has demonstrated to be fully capable of simulating the turbulent flows on unstructured hybrid mesh, as indicated by various comparison tests. The present IGKS solver can be extended to the 3-D case by following a parallel algorithm of LU-SGS in 3D problems reported by Gong et al (2016)~\citep{gong2016an} and considering the neighborhood connectivity of cells carefully, and the relevant work will be reported in the future. As such, the present IGKS solver developed in this study looks promising in its extensive applications.

\section*{Acknowledgements}
The project has been financially supported by the National Natural Science Foundation of China (Grant No. 11472219), the Natural Science Basic Research Plan in Shaanxi Province of China (Program No. 2015JM1002), the 111 Project of China (B17037), the ATCFD Project (2015-F-016), as well as the National Pre-Research Foundation of China.


\bibliographystyle{elsarticle-num}
\bibliography{ImplicitGKS}




\clearpage
\renewcommand\thefigure{\arabic{table}}

\begin{table}[!htbp]
\caption{Vortex center: Vorticity, stream function and location in cavity flow.}\label{table:Table1}
\centering
\begin{tabular}{l|cccccccccc}
\hline
\multirow{2}*{} &	\multicolumn{4}{c}{Primary vortex} & \multicolumn{3}{c}{Left secondary vortex} & \multicolumn{3}{c}{Right secondary vortex} \\
\cline{2-11}
 & $\omega$ & $\psi$ & $x$ & $y$ & $\psi \times {10^4}$ & $x$ & $y$ &  $\psi \times {10^3}$ & $x$ & $y$  \\
\hline
Structured grid & 2.0861 & -0.1178 & 0.5335 & 0.5681 & 2.2461 & 0.0820 & 0.0716 & 1.7123 & 0.8720 & 0.1132 \\
Error (structured grid) & 1.7\% & 0.1\% & 0.4\% & 1.0\% & 2.8\% & 4.5\% & 8.3\% & 2.2\% & 1.5\% & 3.5\% \\
Unstructured mesh & 2.0902 & -0.1179 & 0.5405 & 0.5721 & 2.2870 & 0.0899 & 0.0862 & 1.7289 & 0.8752 & 0.1145 \\
Error (Unstructured grid) & 2.0\% & 0.0\% & 1.7\% & 1.7\% & 1.1\% & 4.7\% & 10.4\% & 1.3\% & 1.8\% & 4.7\% \\
Hou et al~\cite{hou1995simulation} & 2.0760 & -0.1178 & 0.5333 & 0.5647 & 2.2200 & 0.0902 & 0.0784 & 1.6900 & 0.8667 & 0.1137 \\
Zhuo et al~\cite{zhuo2012filter} & 2.0570 & -0.1179 & 0.5311 & 0.5662 & 2.2667 & 0.0828 & 0.0770 & 1.7066 & 0.8645 & 0.1120 \\
Ghia et al~\cite{ghia1982high} & 2.0497 & -0.1179 & 0.5313 & 0.5625 & 2.3113 & 0.0859 & 0.0781 & 1.7510 & 0.8594 & 0.1094 \\
\hline
\end{tabular}\\
\end{table}

\begin{table}[!htbp]
\caption{Comparison of computation time for explicit and implicit method.}\label{table:Table2}
\centering
\begin{tabular}{c|c|cccc|c}
\hline
\multirow{2}*{} & \multirow{2}*{Number of cells}  & \multicolumn{2}{c}{Explicit method} &  \multicolumn{2}{c|}{Implicit method} & \multirow{2}*{Speedup} \\
\cline{3-6}
 &  & Iteration steps & Time(s) & Iteration steps & Time(s) \\
\hline
Cavity (structured grid)  & 10000 & 608000 & 34820.2 & 4430 & 543.2 & 64.1\\
Cavity (unstructured mesh) & 8933  & 590000 & 30184.0 & 4270 & 484.4 & 62.3\\
Laminar flat plate  & 13441 & 605000 & 64685.6 & 4320 & 717.4 & 90.2\\
Turbulent flat plate  & 26386 & 810000 & 78599.5 & 5340 & 1087.7 & 72.3\\
NHLP multi-element airfoil  & 64344 & 520000 & 123614.2 & 3650 & 1308.4 & 94.5\\
\hline
\end{tabular}\\
\end{table}

\clearpage
\renewcommand\thefigure{\arabic{figure}}

\begin{figure}
  \centering
  \includegraphics[width=0.5\textwidth]{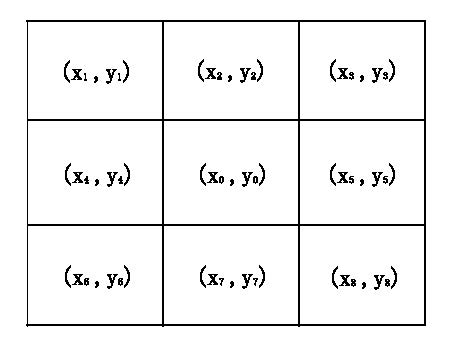}
  \caption{Cells for reconstruction.}
  \label{fig:Fig01}
\end{figure}

\begin{figure}
  \centering
  \includegraphics[width=0.6\textwidth]{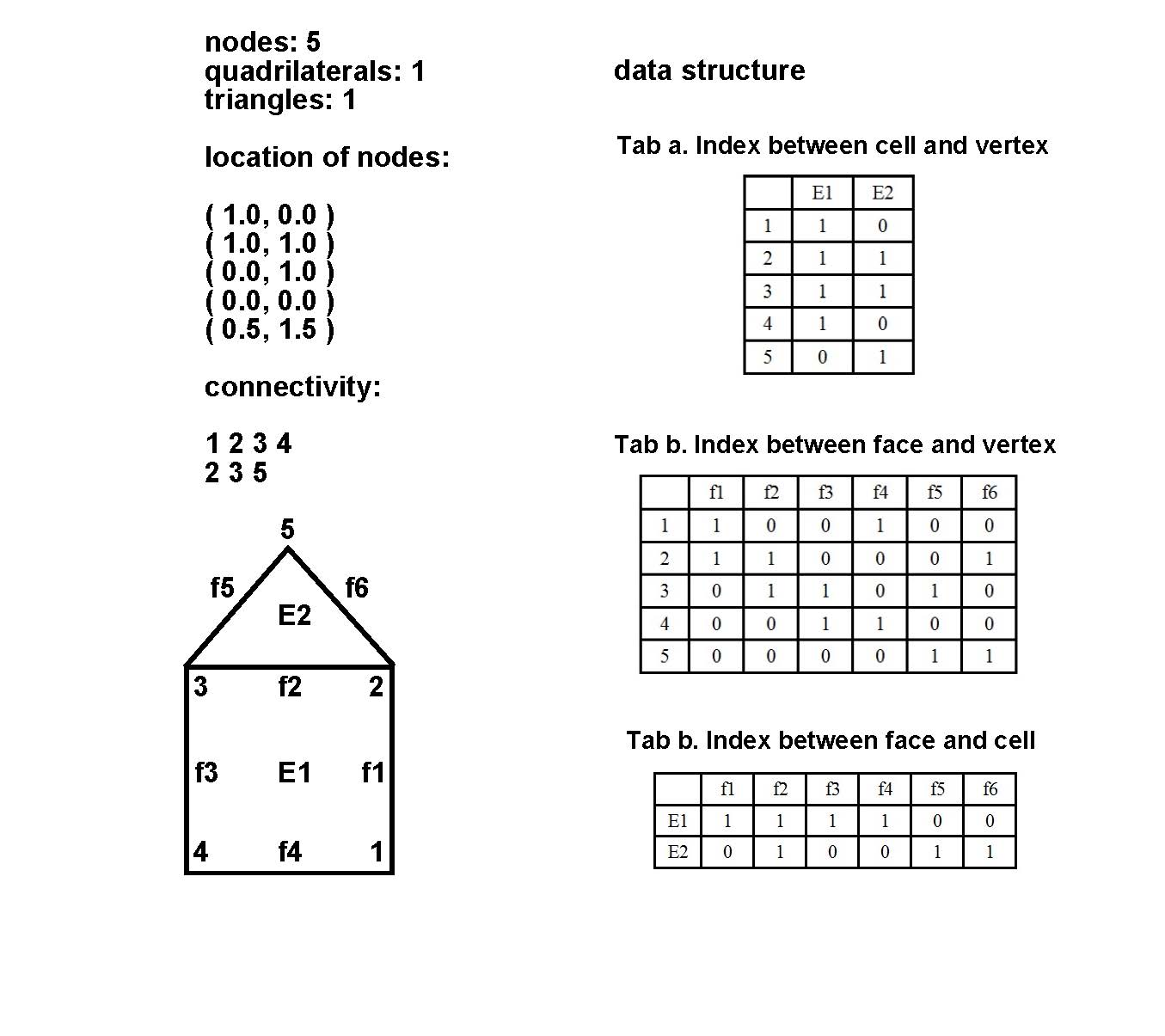}
  \caption{An example of NHMD.}
  \label{fig:Fig02}
\end{figure}

\begin{figure}
  \centering
  \includegraphics[width=0.5\textwidth]{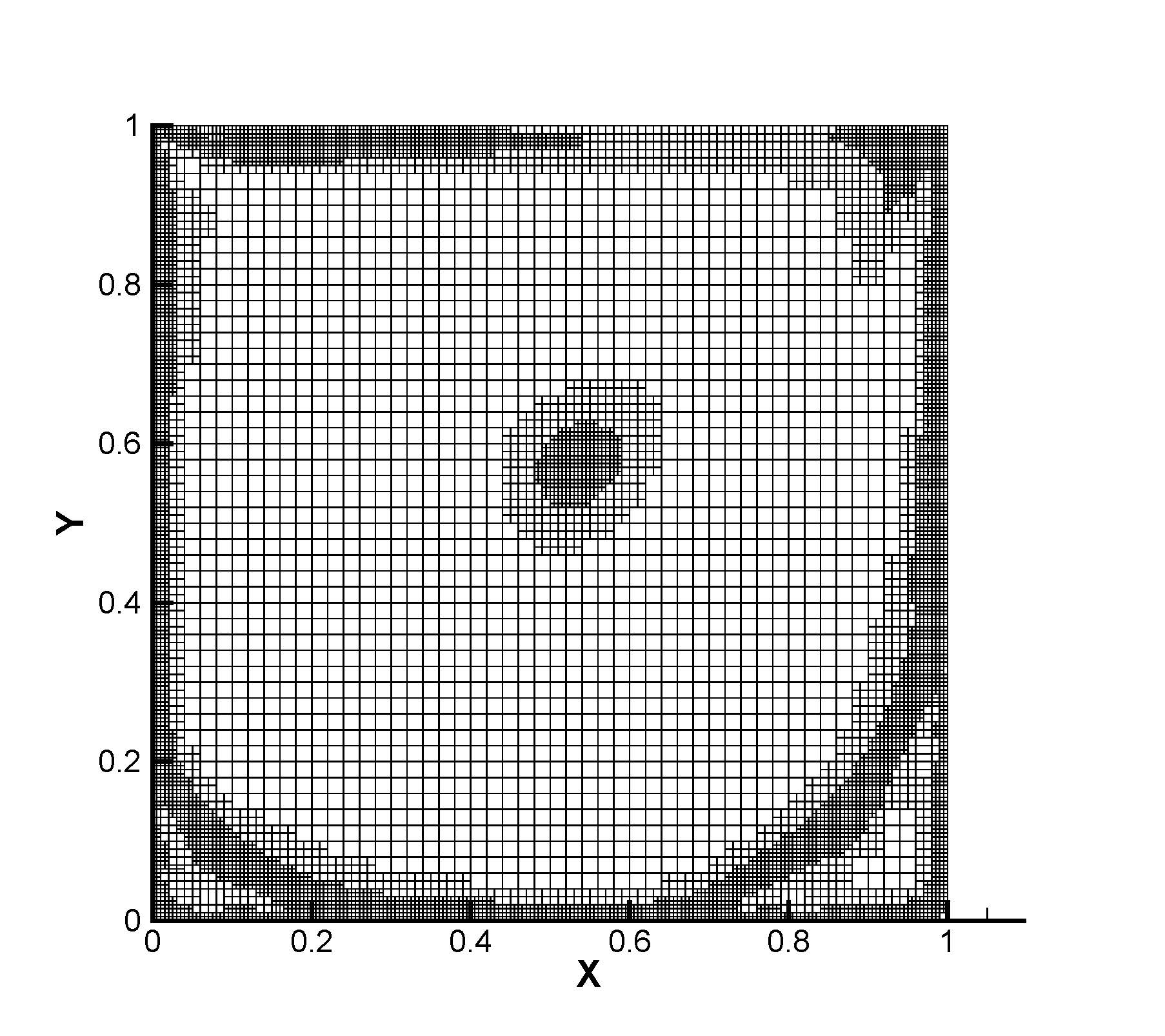}
  \caption{The unstructured hybrid mesh for cavity flow simulation.}
  \label{fig:Fig03}
\end{figure}

\begin{figure}
  \centering
  \includegraphics[width=0.5\textwidth]{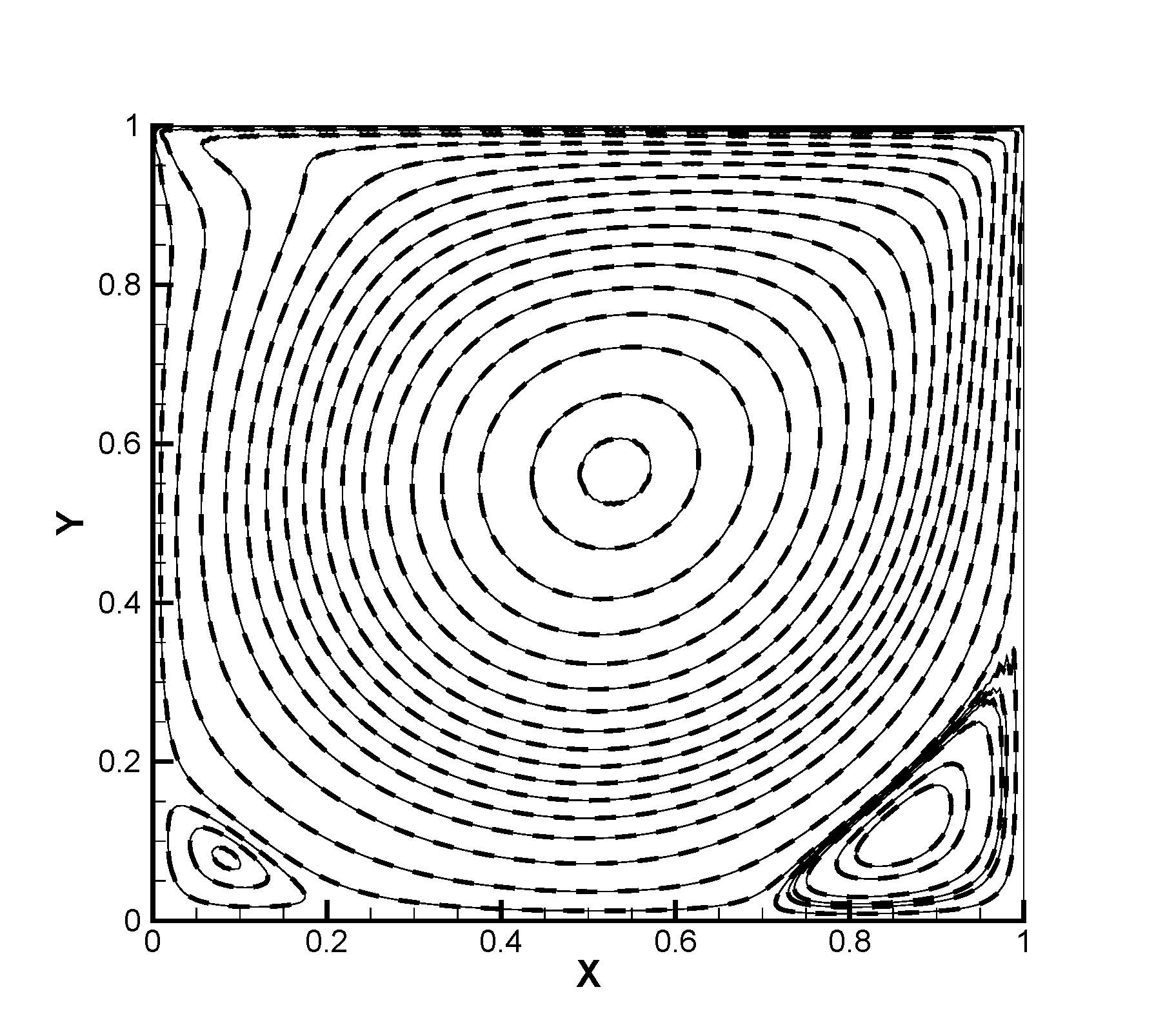}
  \caption{Contours of the stream function in lid-driven cavity flow on uniform mesh (solid line) and unstructured hybrid mesh (dashed line).}
  \label{fig:Fig04}
\end{figure}

\begin{figure}
  \centering
  \includegraphics[width=0.5\textwidth]{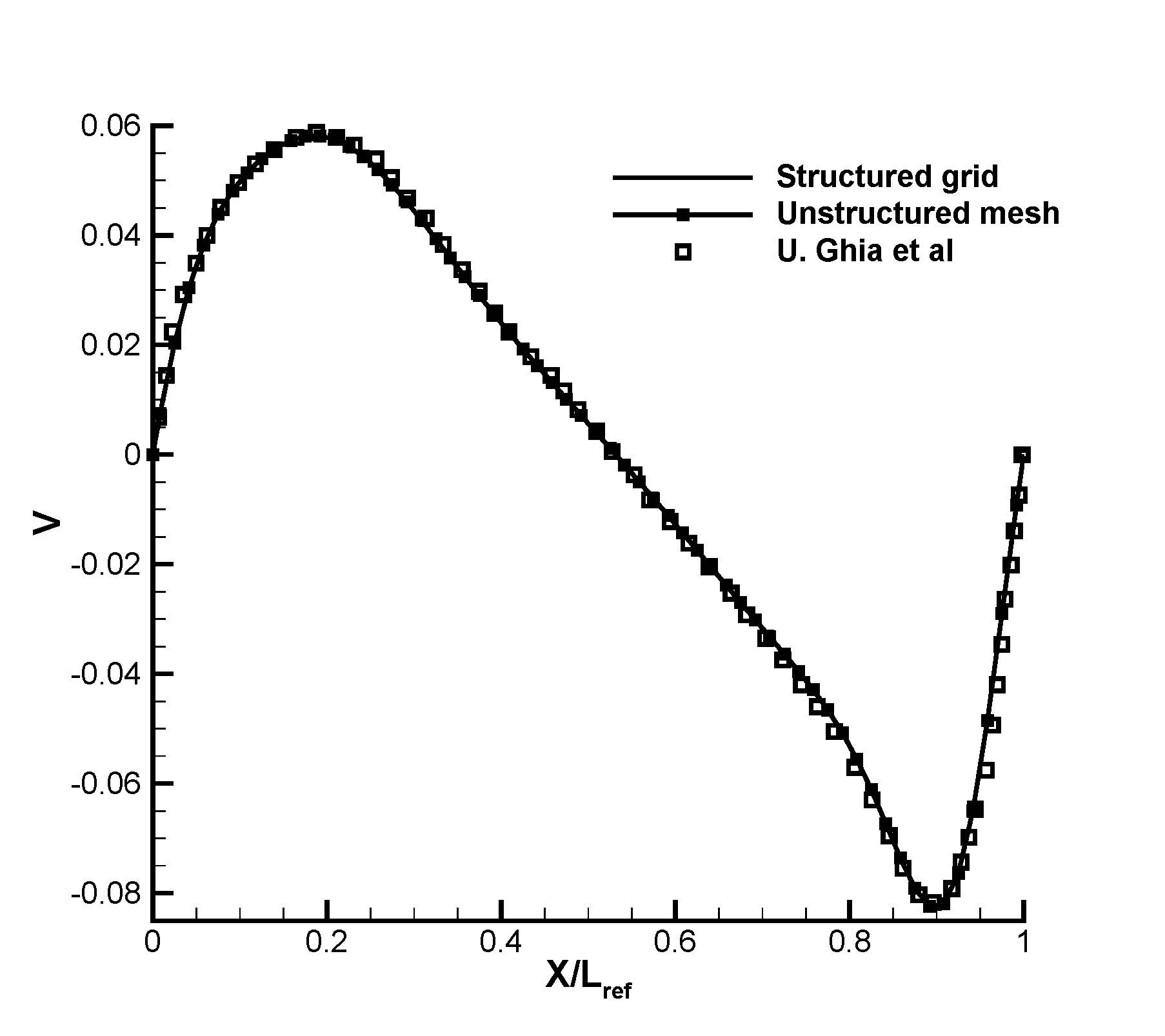}
  \caption{Vertical velocity pofile along the mid-height line in cavity.}
  \label{fig:Fig05}
\end{figure}

\begin{figure}
  \centering
  \includegraphics[width=0.5\textwidth]{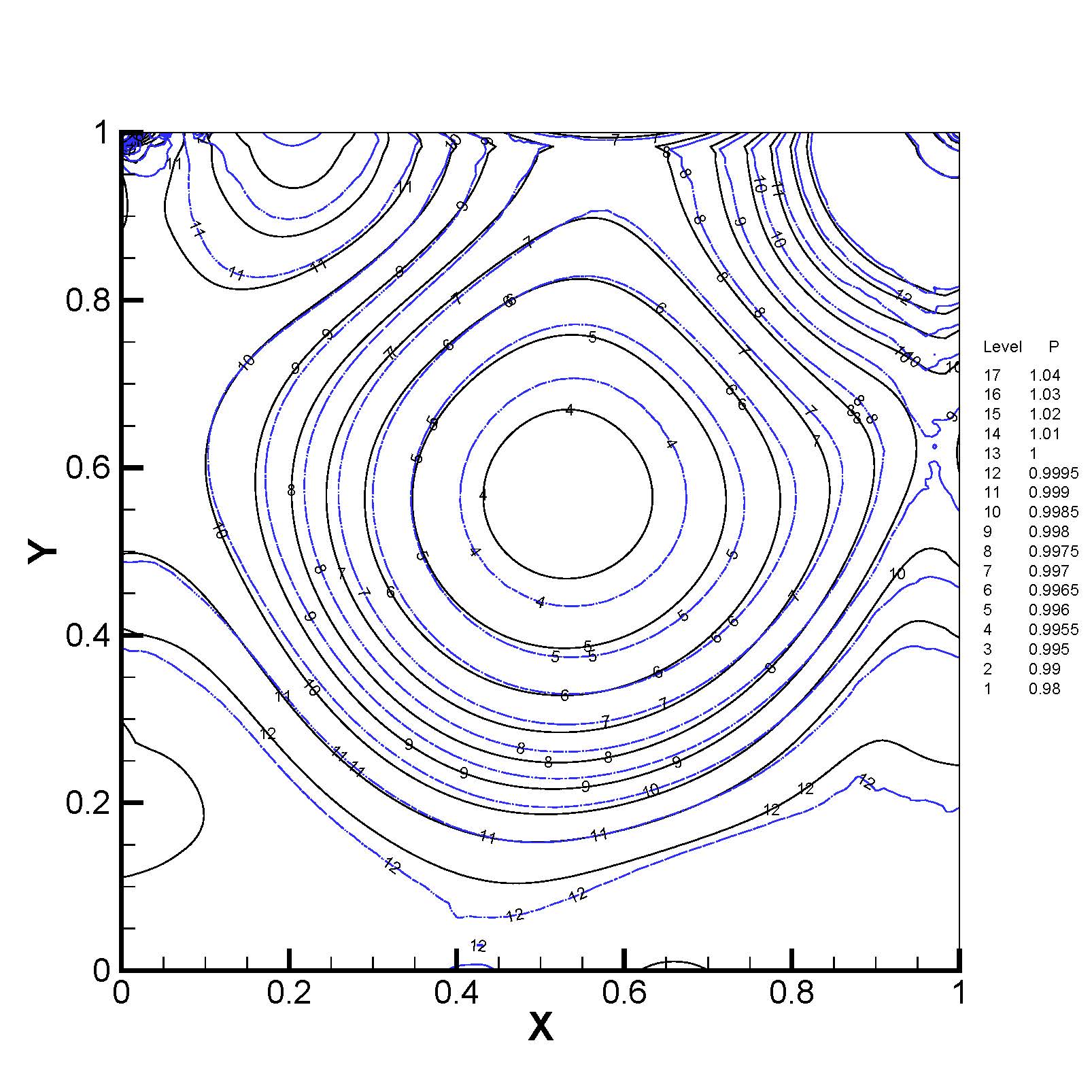}
  \caption{Pressure contours of lid-driven cavity flow by implicit GKS on uniform mesh (solid lines) and unstructured hybrid mesh (dash-dotted lines).}
  \label{fig:Fig06}
\end{figure}

\begin{figure}
  \centering
  \includegraphics[width=0.5\textwidth]{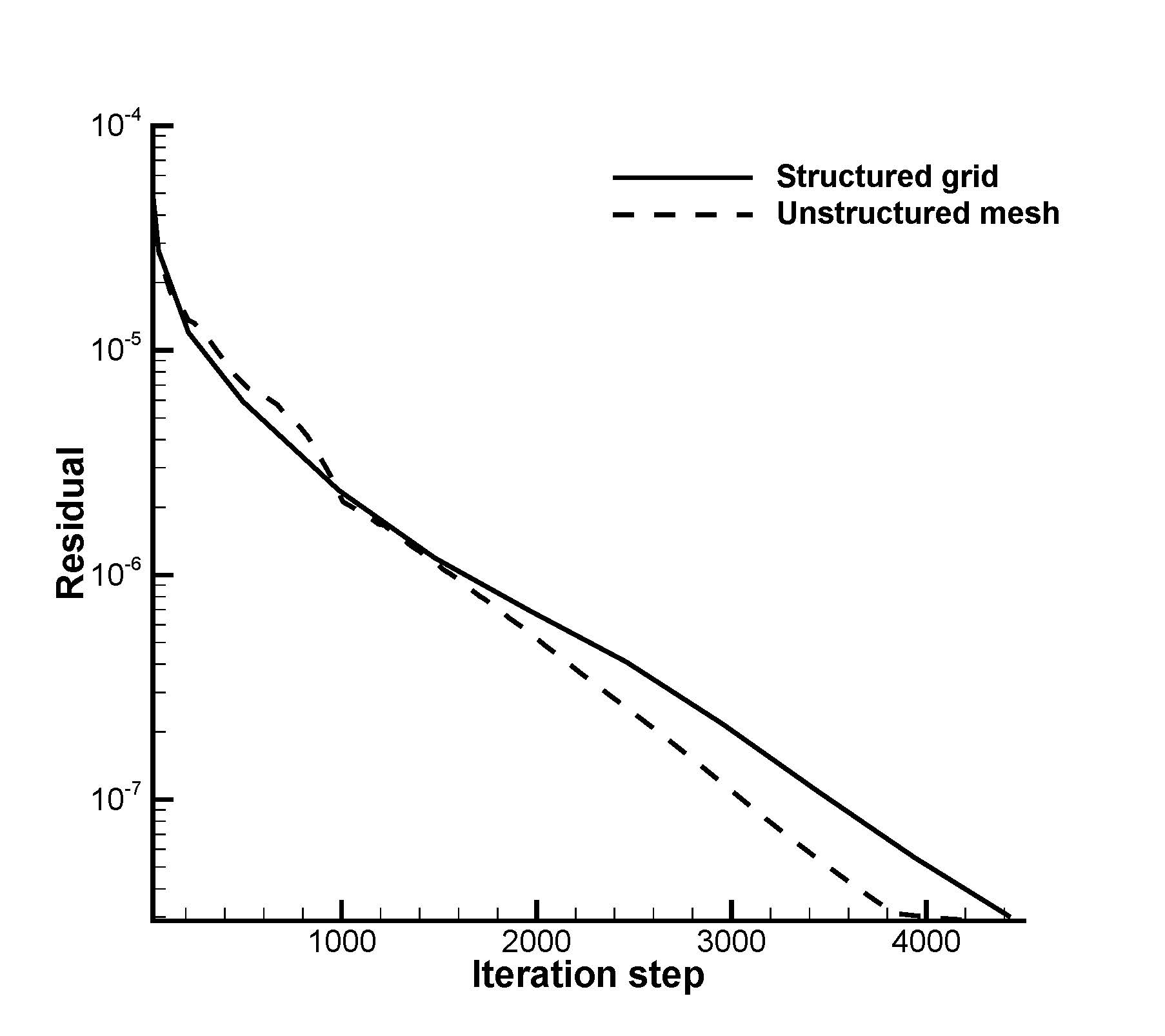}
  \caption{The residual curves of lid-driven cavity flow by implicit GKS on uniform mesh and unstructured hybrid mesh.}
  \label{fig:Fig07}
\end{figure}

\begin{figure}
  \centering
  \includegraphics[width=0.5\textwidth]{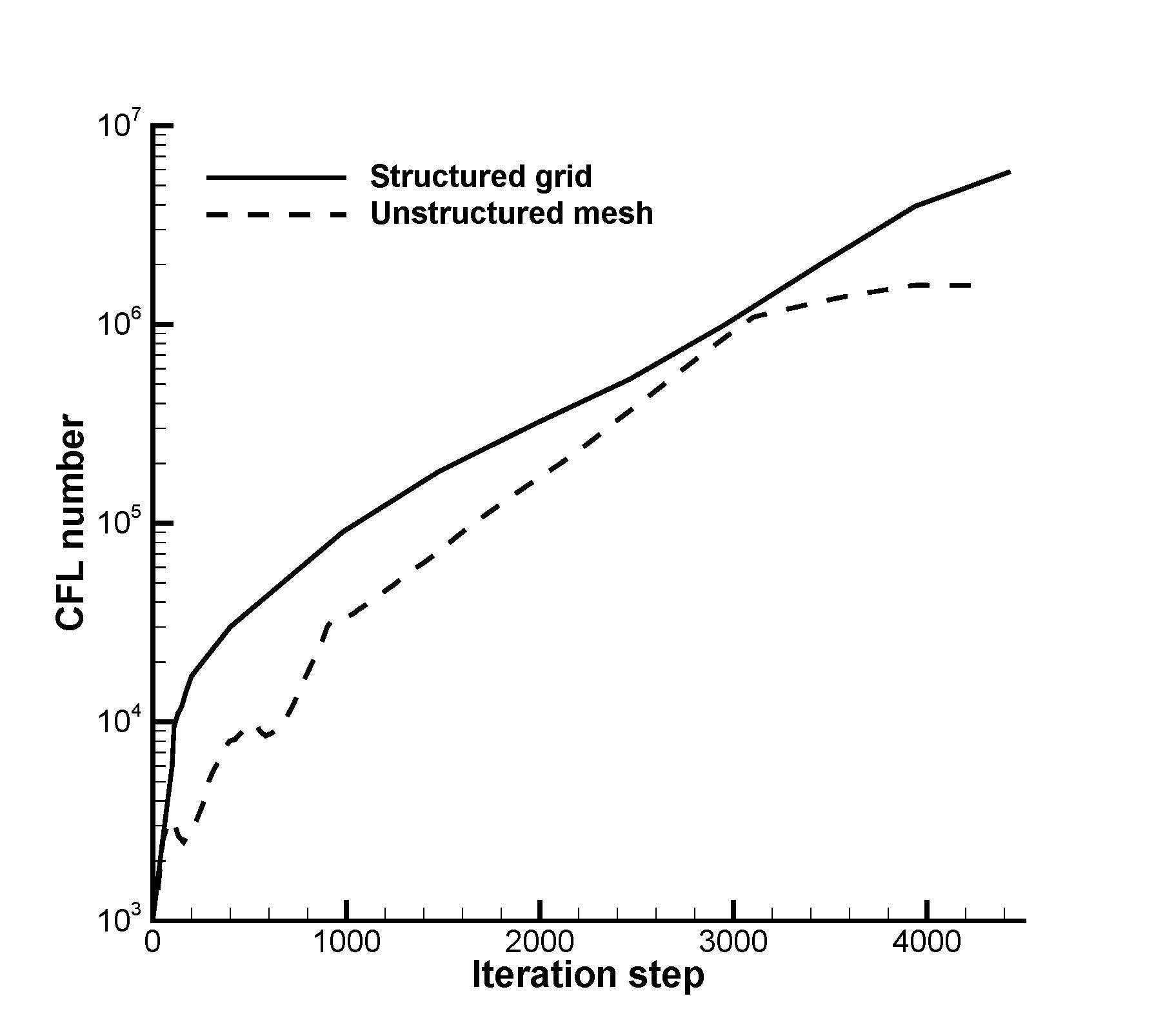}
  \caption{The CFL number curve of lid-driven cavity flow on uniform mesh and unstructured hybrid mesh.}
  \label{fig:Fig08}
\end{figure}

\begin{figure}
  \centering
  \includegraphics[width=0.5\textwidth]{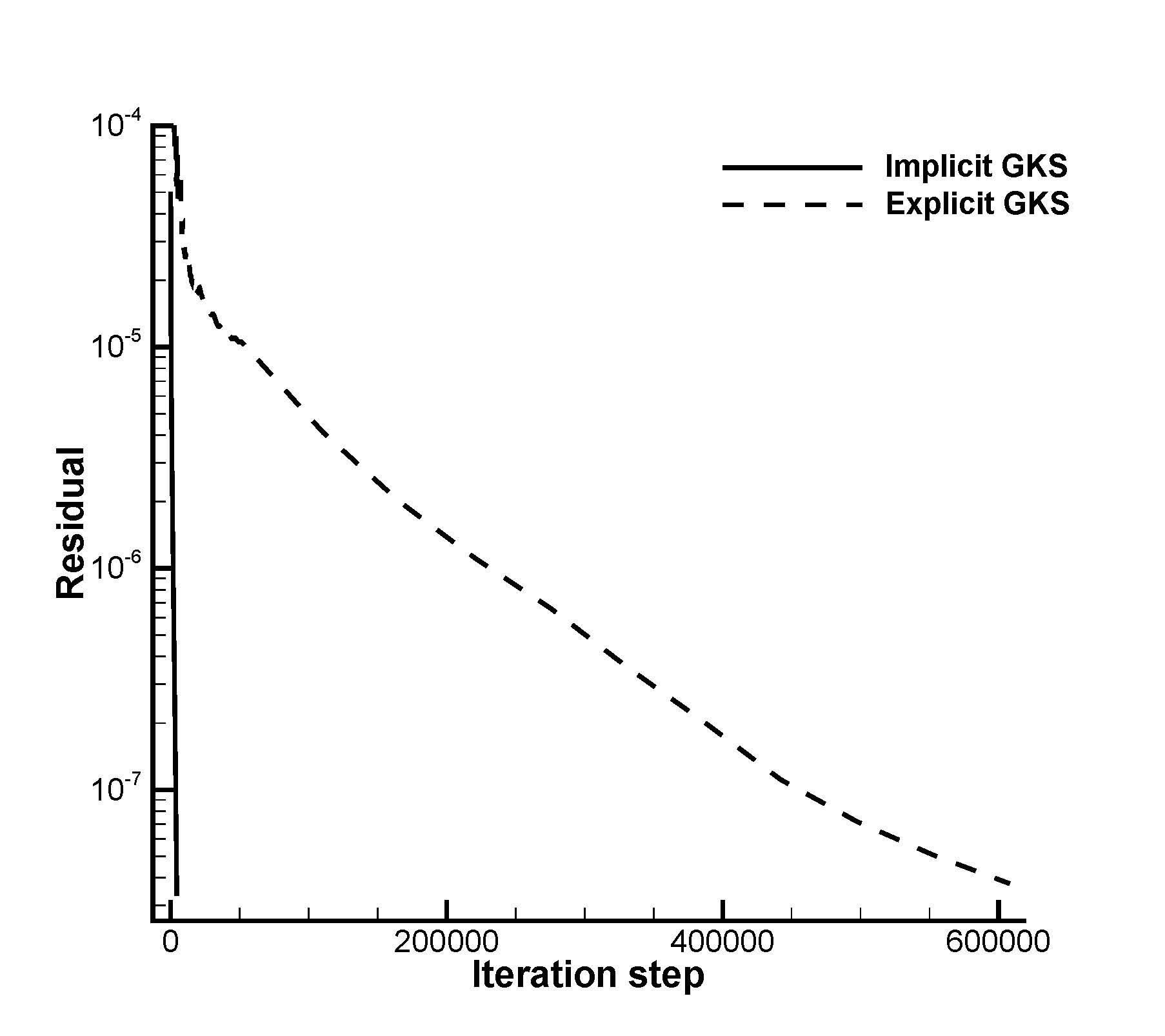}
  \caption{The residual of lid-driven cavity flow on uniform mesh.}
  \label{fig:Fig09}
\end{figure}

\begin{figure}
  \centering
  \includegraphics[width=0.5\textwidth]{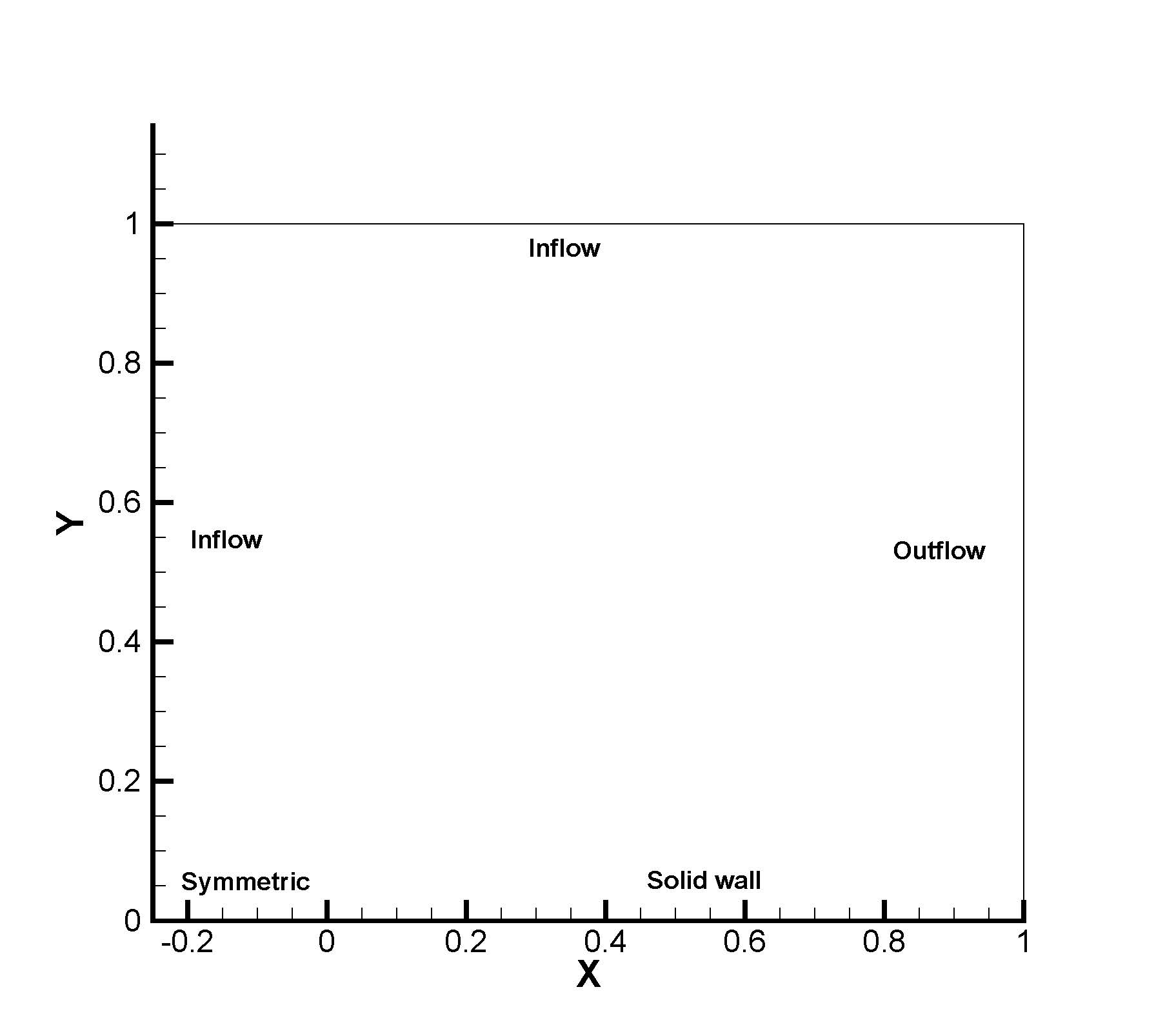}
  \caption{The computational domain for incompressible laminar flow over a flat plate.}
  \label{fig:Fig10}
\end{figure}

\clearpage

\begin{figure}
  \centering
  \subfigure[]{\label{fig:Fig11a}\includegraphics[width=0.45\textwidth]{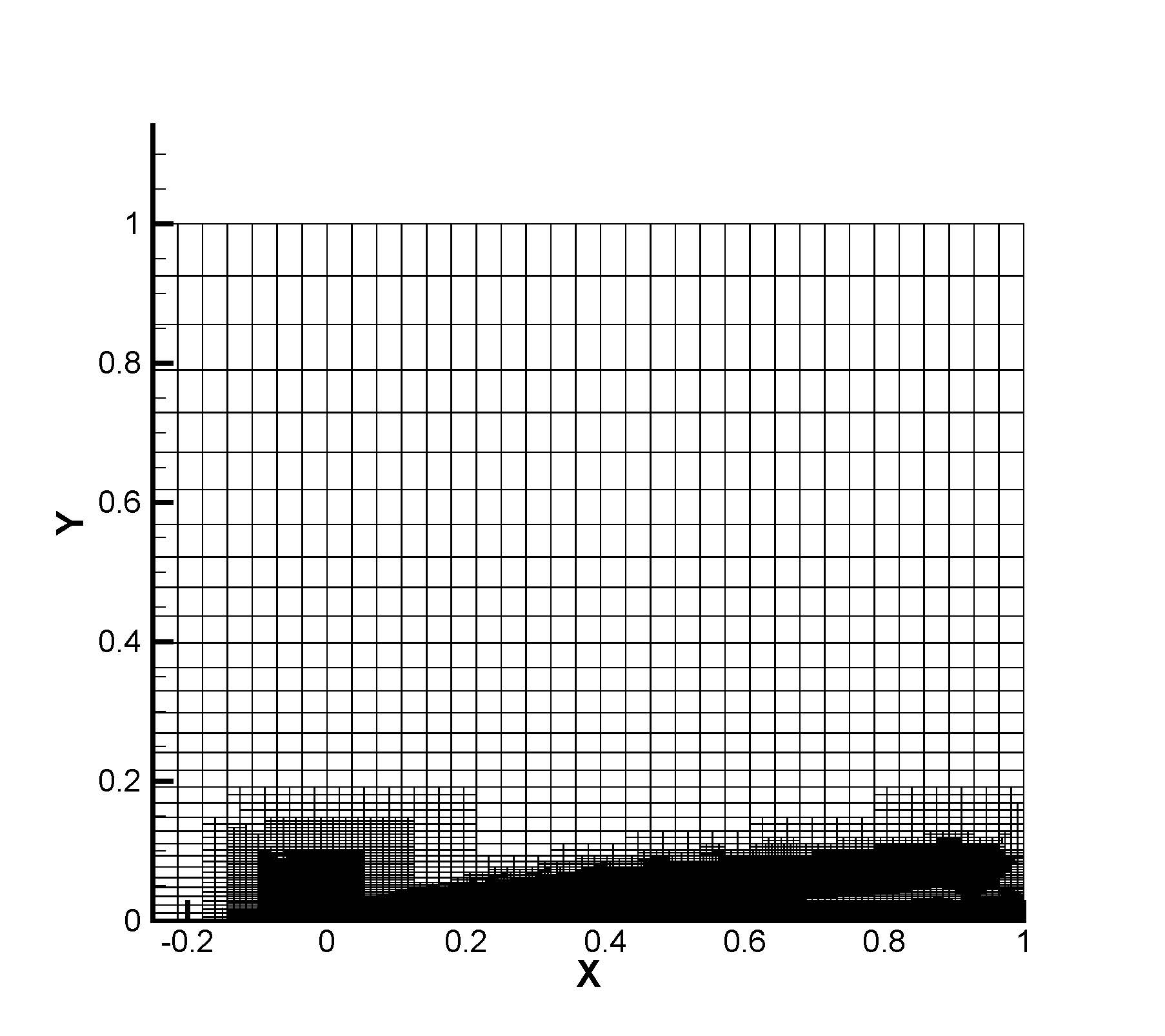}}
  \subfigure[]{\label{fig:Fig11b}\includegraphics[width=0.45\textwidth]{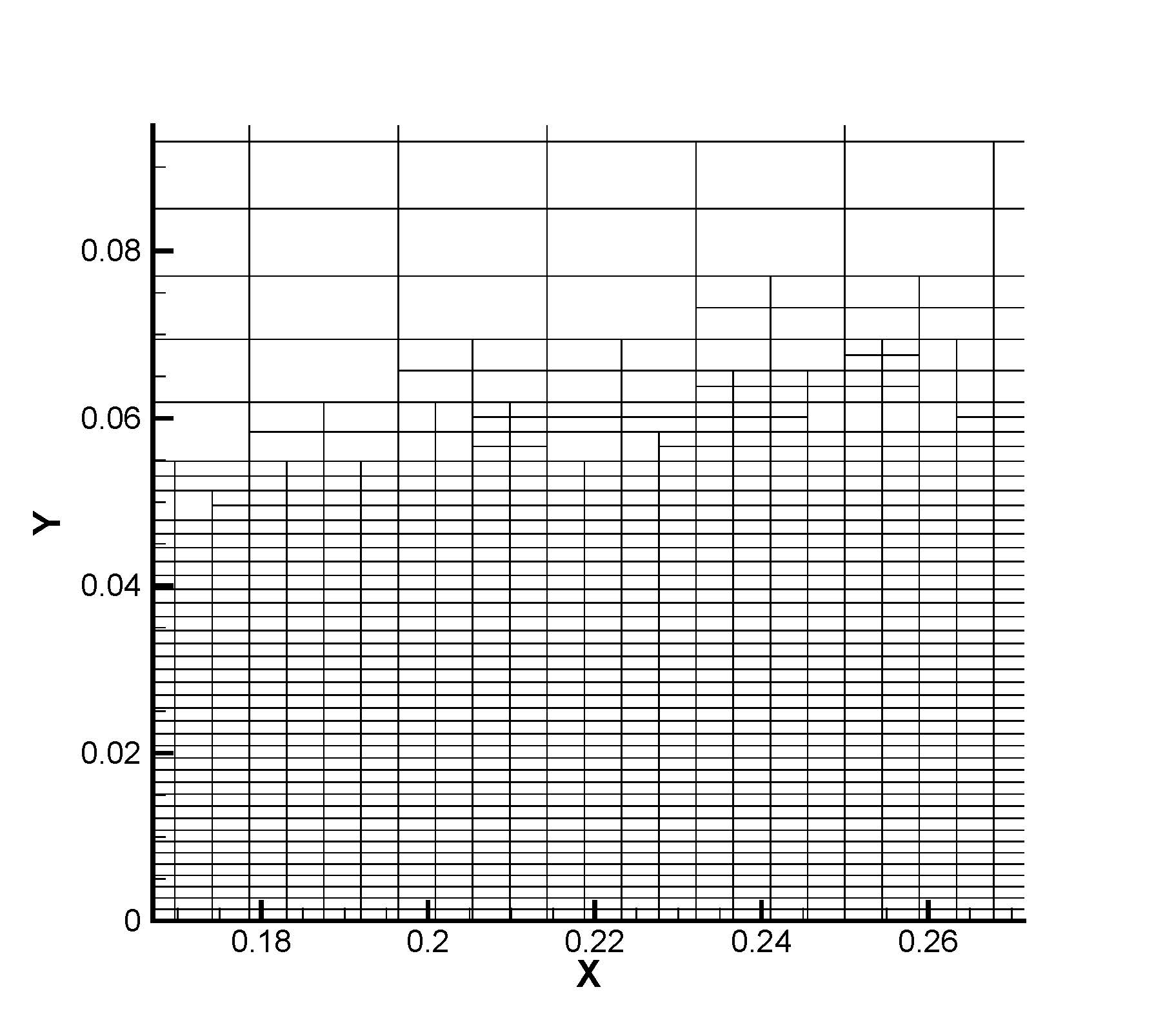}}
  \caption{The hybrid mesh for incompressible laminar flow over a flat plate. \\(a) Hybrid mesh for incompressible laminar flow, (b) Mesh near wall.}
  \label{fig:Fig11}
\end{figure}

\begin{figure}
  \centering
  \subfigure[]{\label{fig:Fig12a}\includegraphics[width=0.40\textwidth]{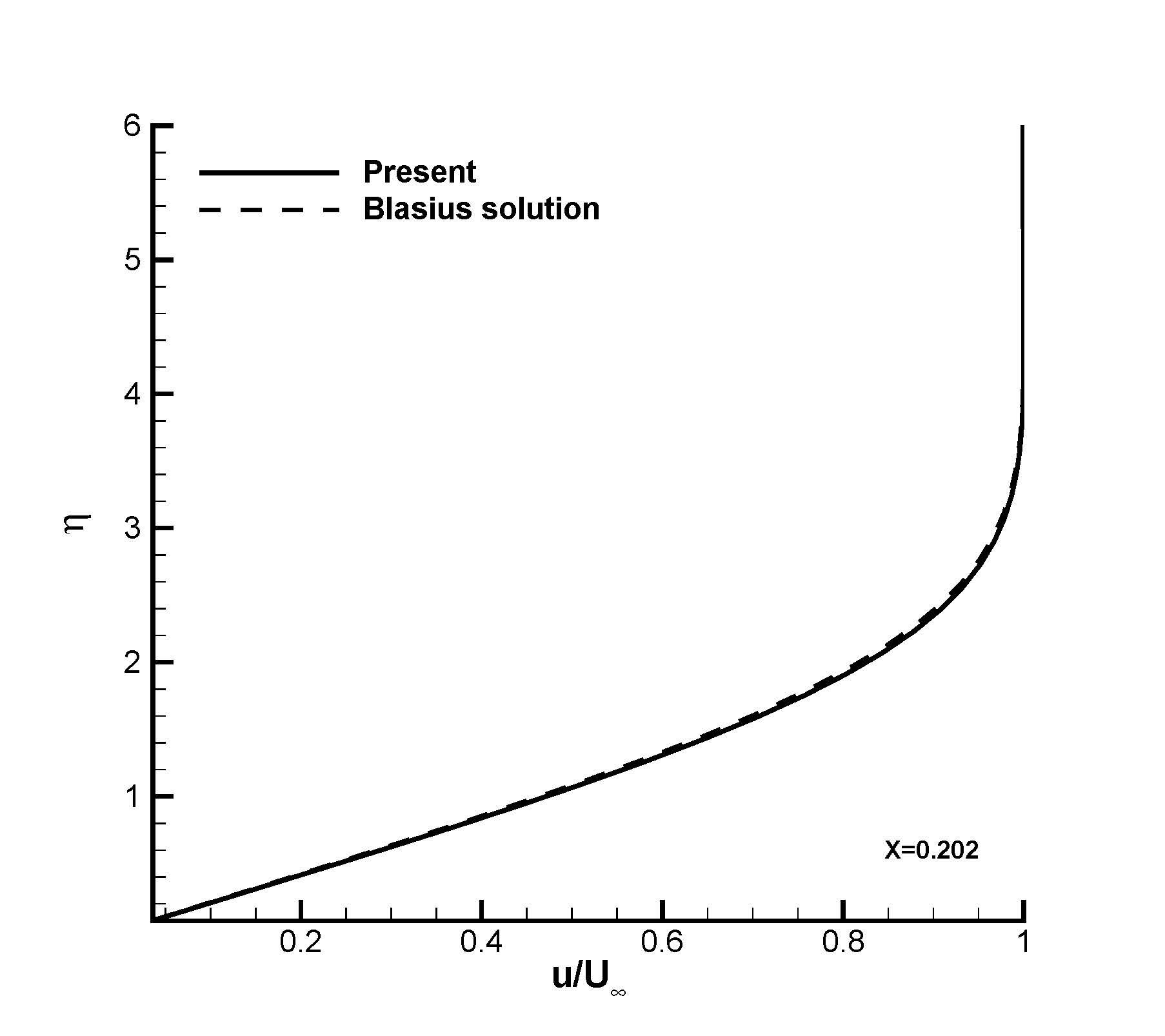}}
  \subfigure[]{\label{fig:Fig12b}\includegraphics[width=0.40\textwidth]{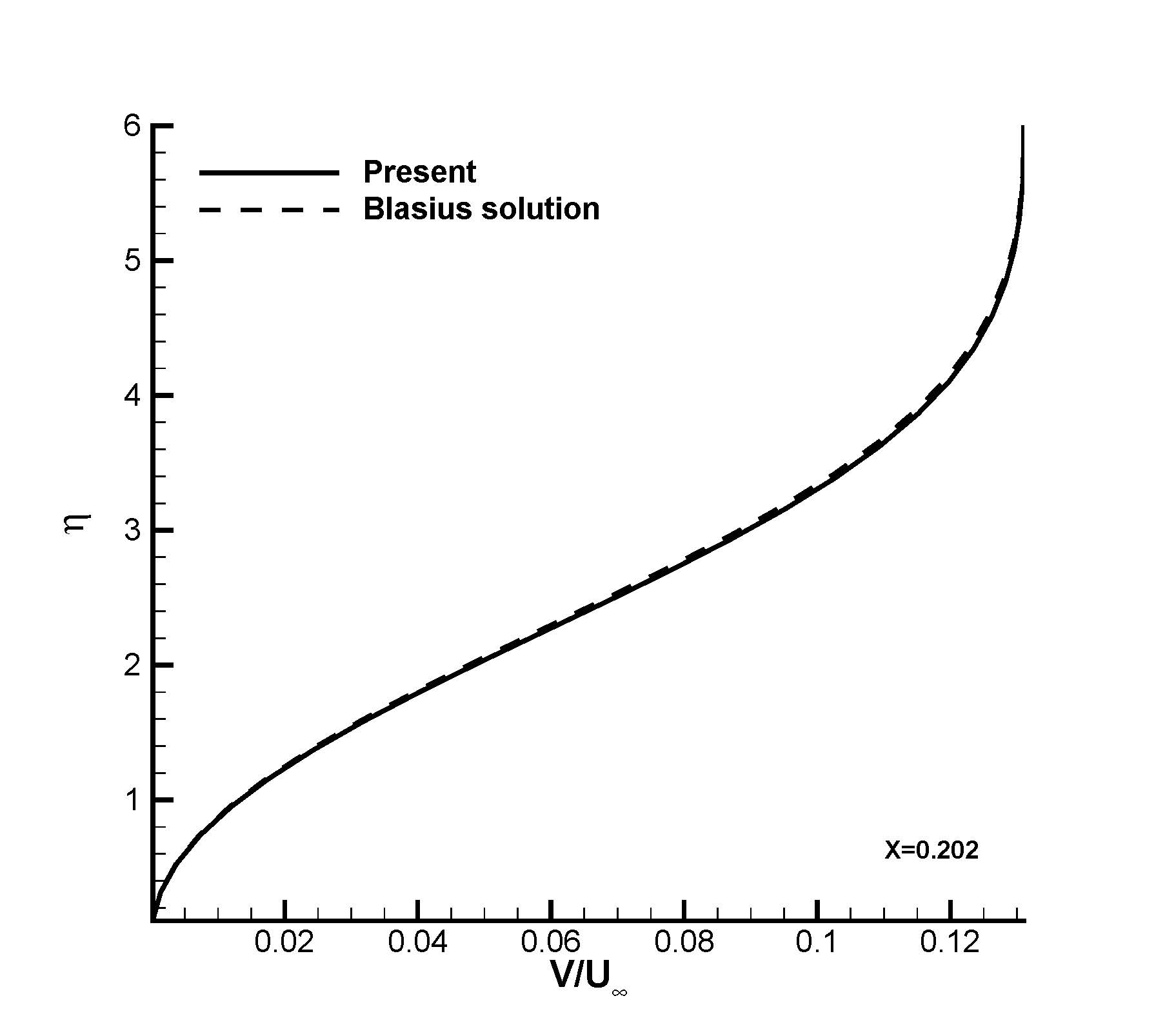}}
  \subfigure[]{\label{fig:Fig12v}\includegraphics[width=0.40\textwidth]{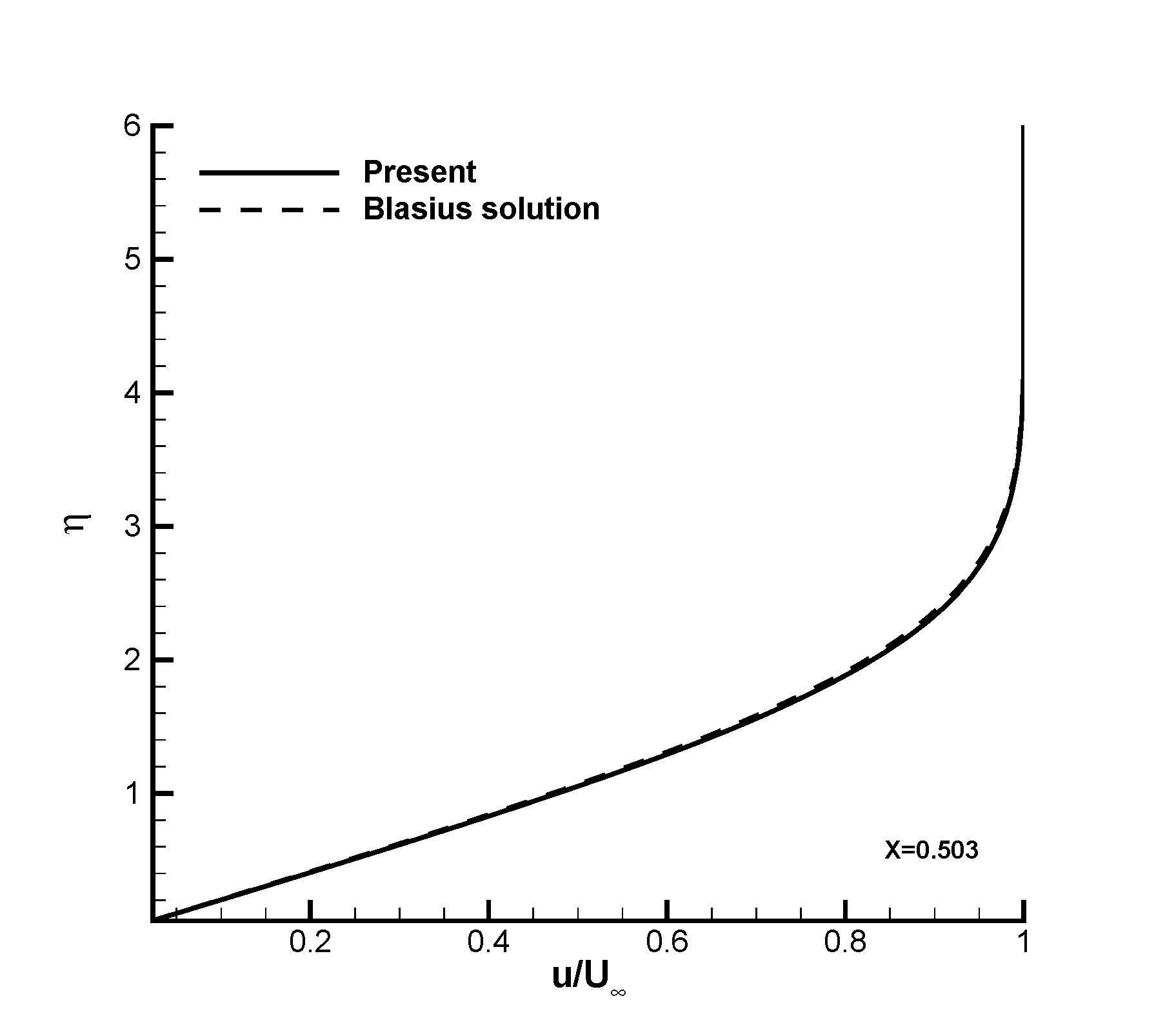}}
  \subfigure[]{\label{fig:Fig12d}\includegraphics[width=0.40\textwidth]{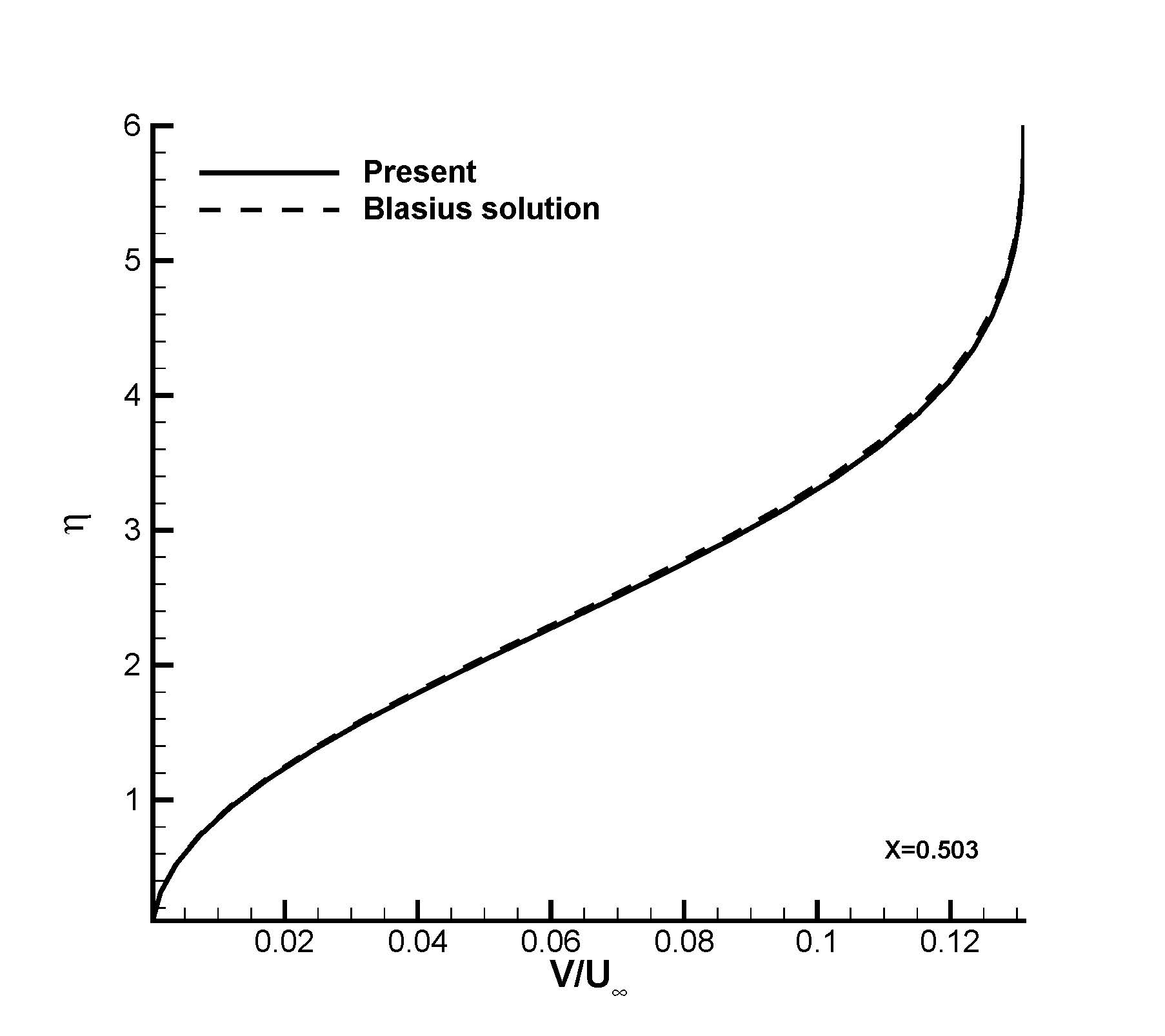}}
  \subfigure[]{\label{fig:Fig12e}\includegraphics[width=0.40\textwidth]{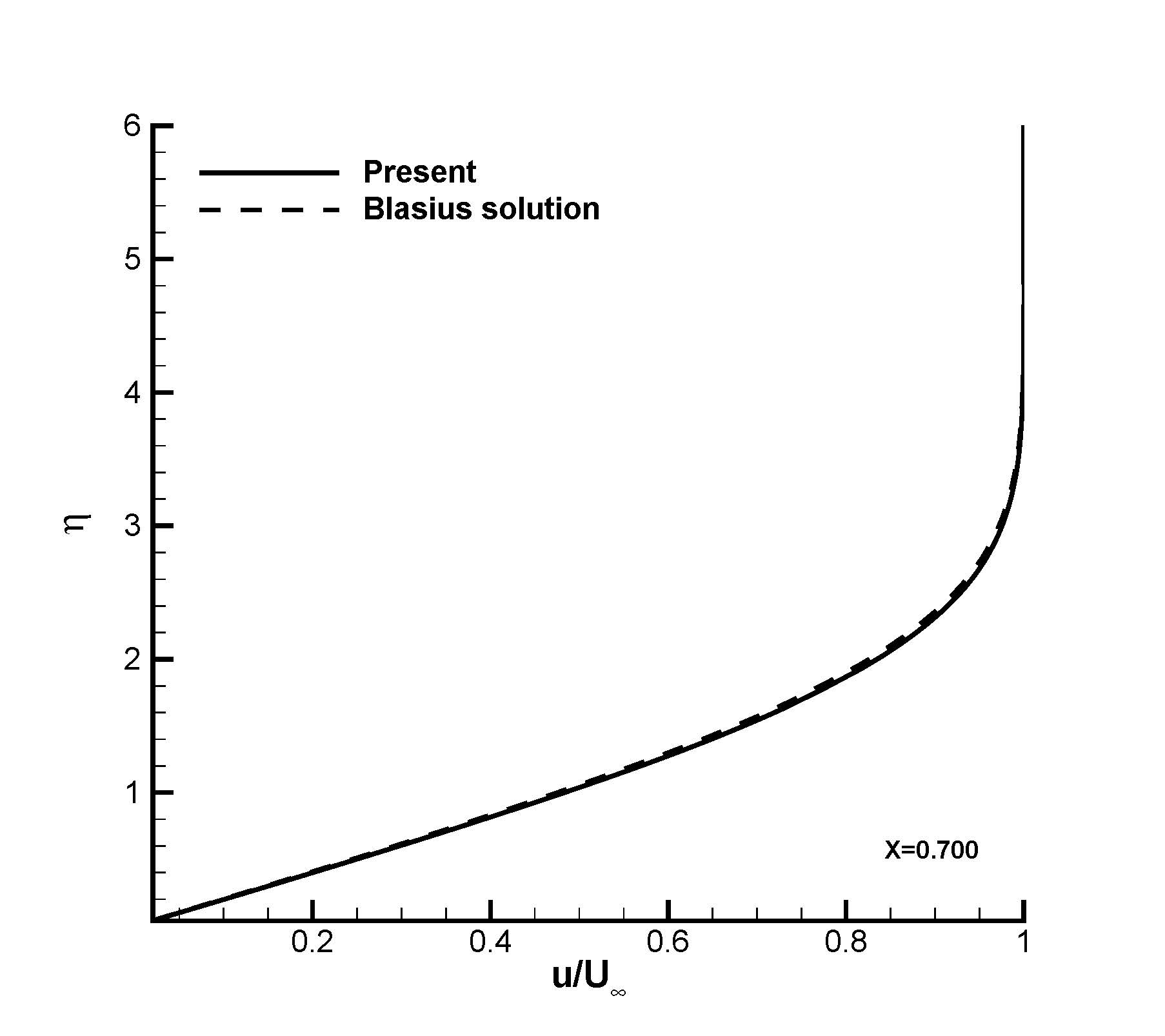}}
  \subfigure[]{\label{fig:Fig12f}\includegraphics[width=0.40\textwidth]{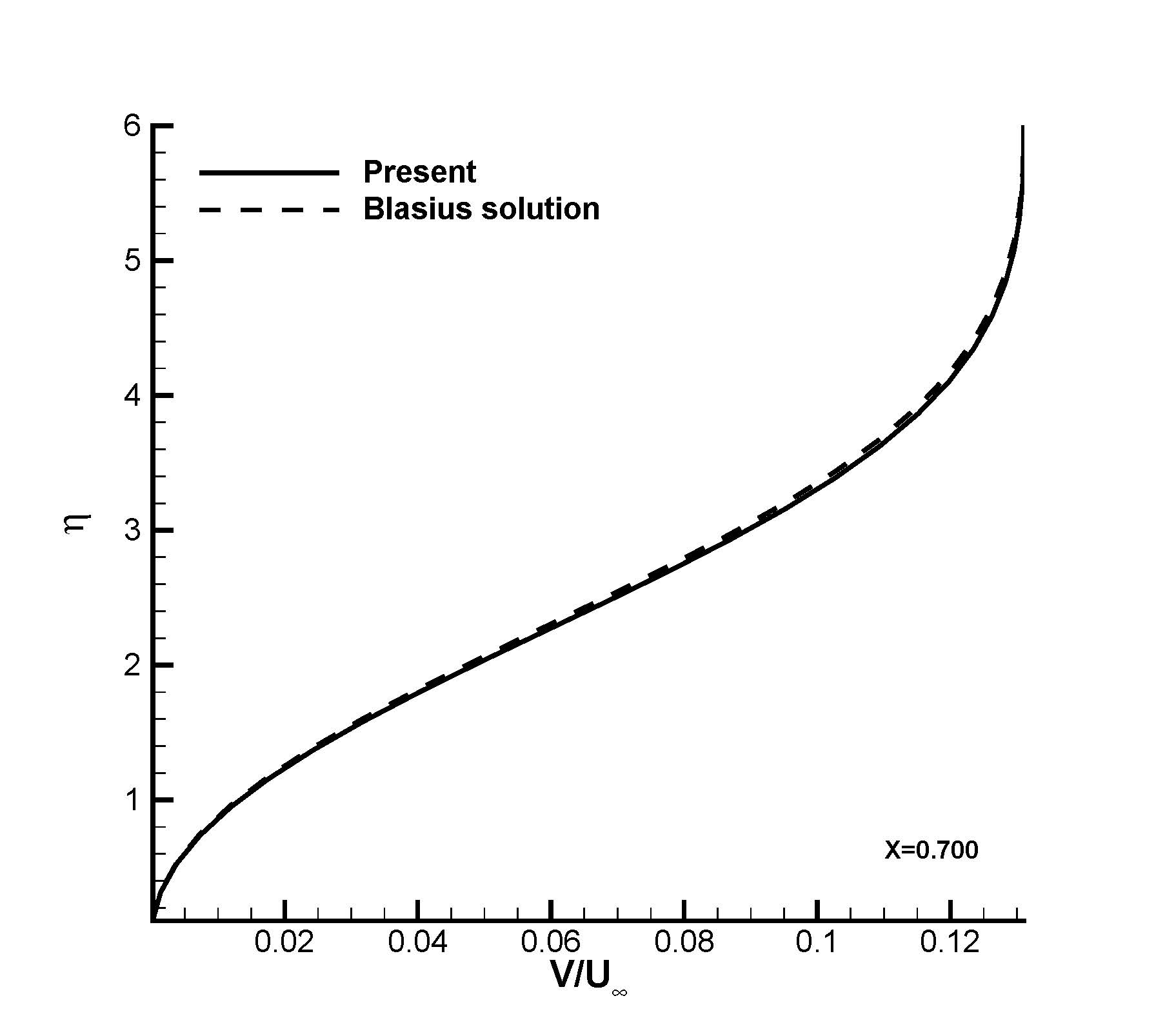}}
  \caption{Velocity profiles of incompressible laminar flow over a flat plate at three X-locations.}
  \label{fig:Fig12}
\end{figure}

\begin{figure}
  \centering
  \includegraphics[width=0.5\textwidth]{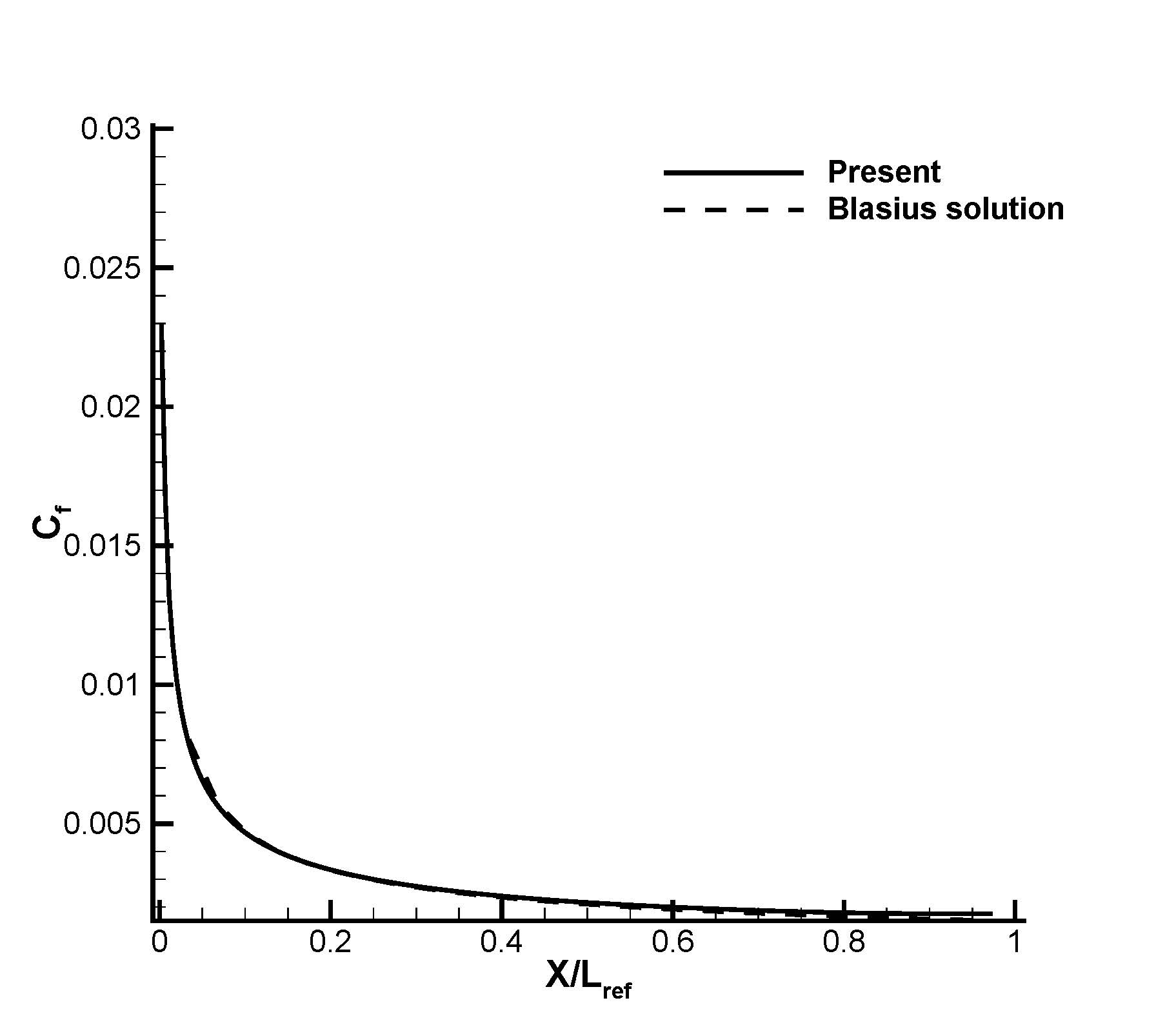}
  \caption{The skin friction coefficient comparison for a incompressible laminar flat plate.}
  \label{fig:Fig13}
\end{figure}

\begin{figure}
  \centering
 \subfigure[]{\label{fig:Fig14a}\includegraphics[width=0.45\textwidth]{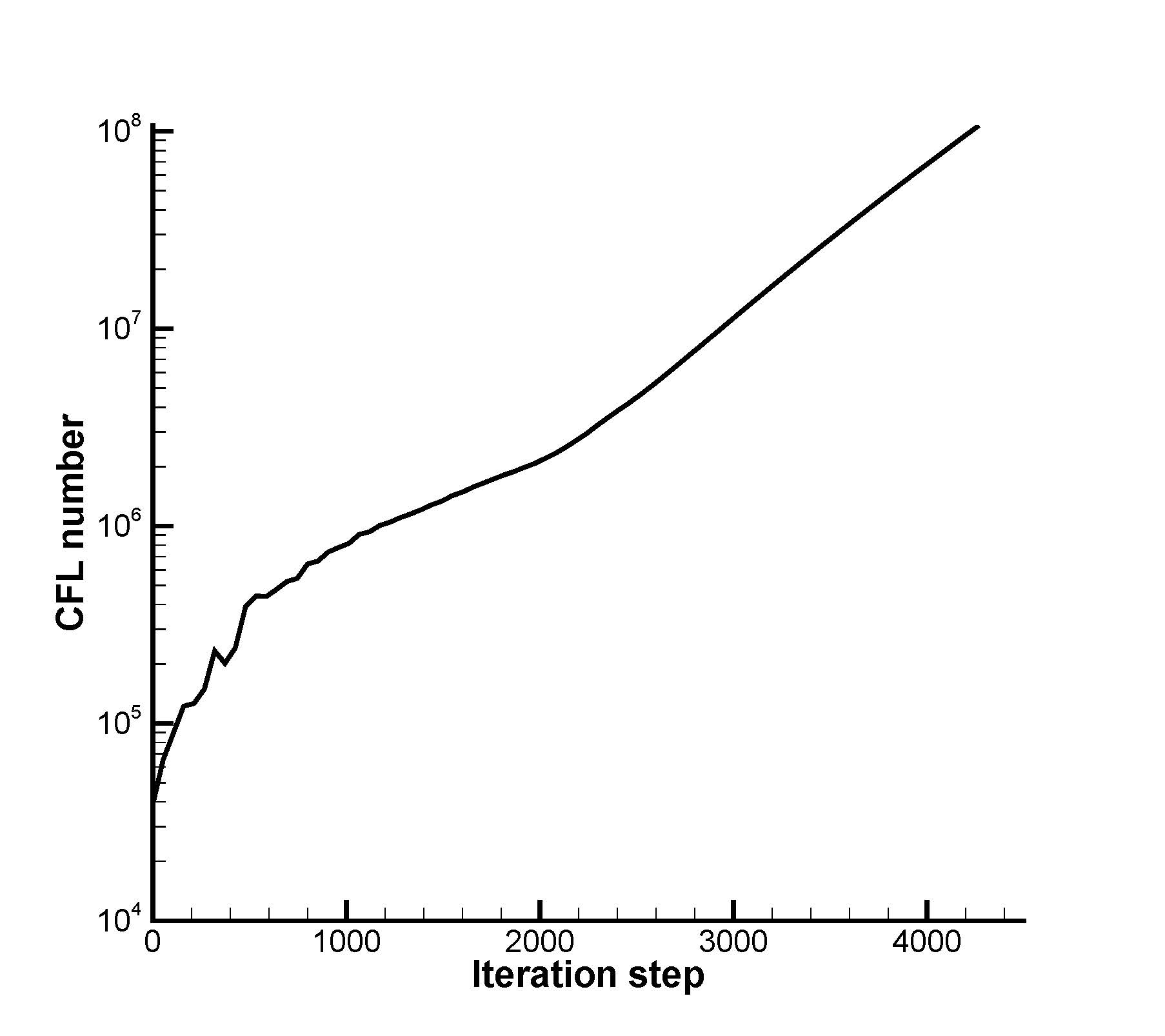}}
 \subfigure[]{\label{fig:Fig14b}\includegraphics[width=0.45\textwidth]{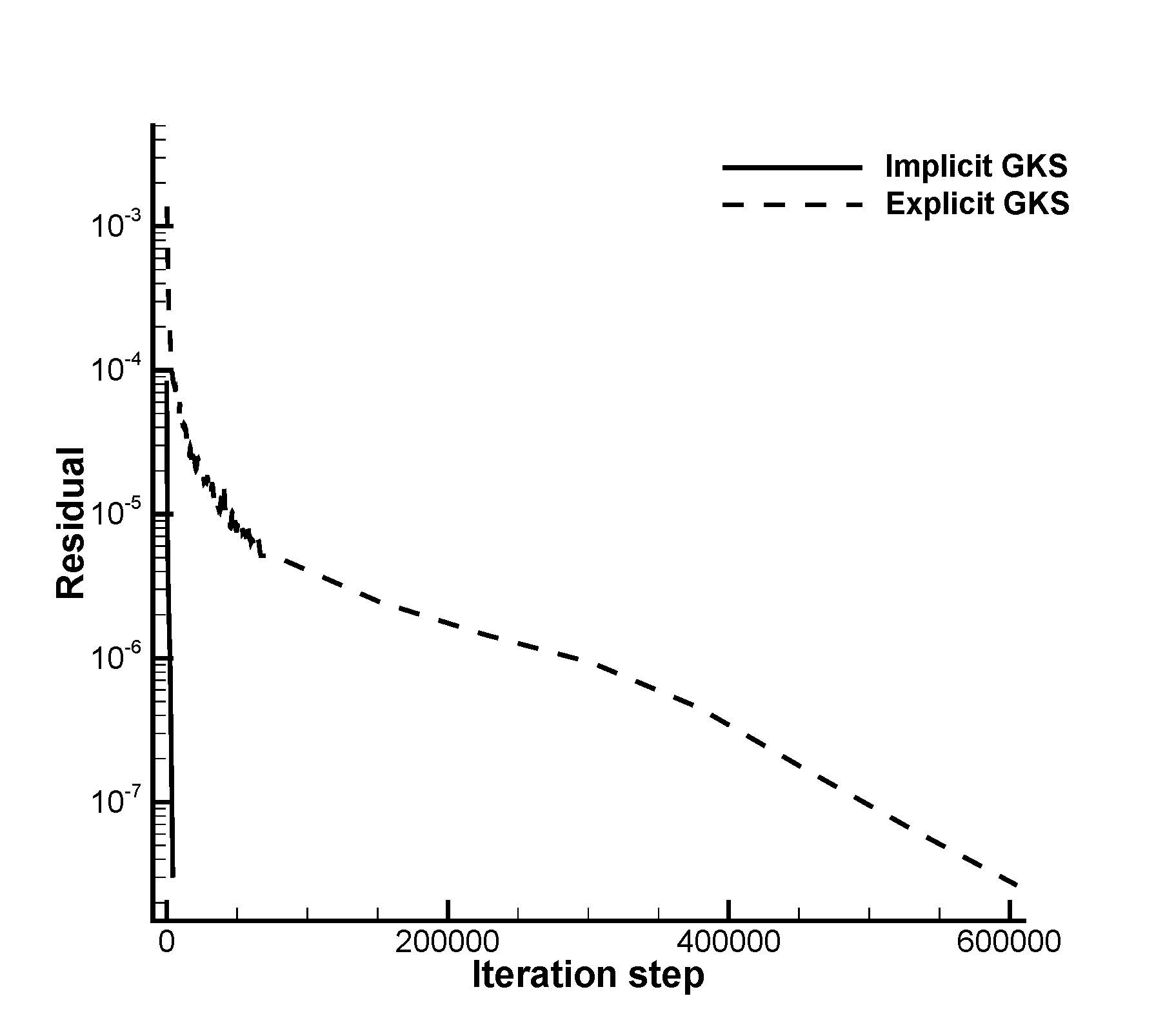}}
  \caption{The CFL number and residual curves of incompressible laminar flow over a flat plate.\\(a) CFL number, (b) Residual. }
  \label{fig:Fig14}
\end{figure}

\begin{figure}
  \centering
  \subfigure[]{\label{fig:Fig15a}\includegraphics[width=0.45\textwidth]{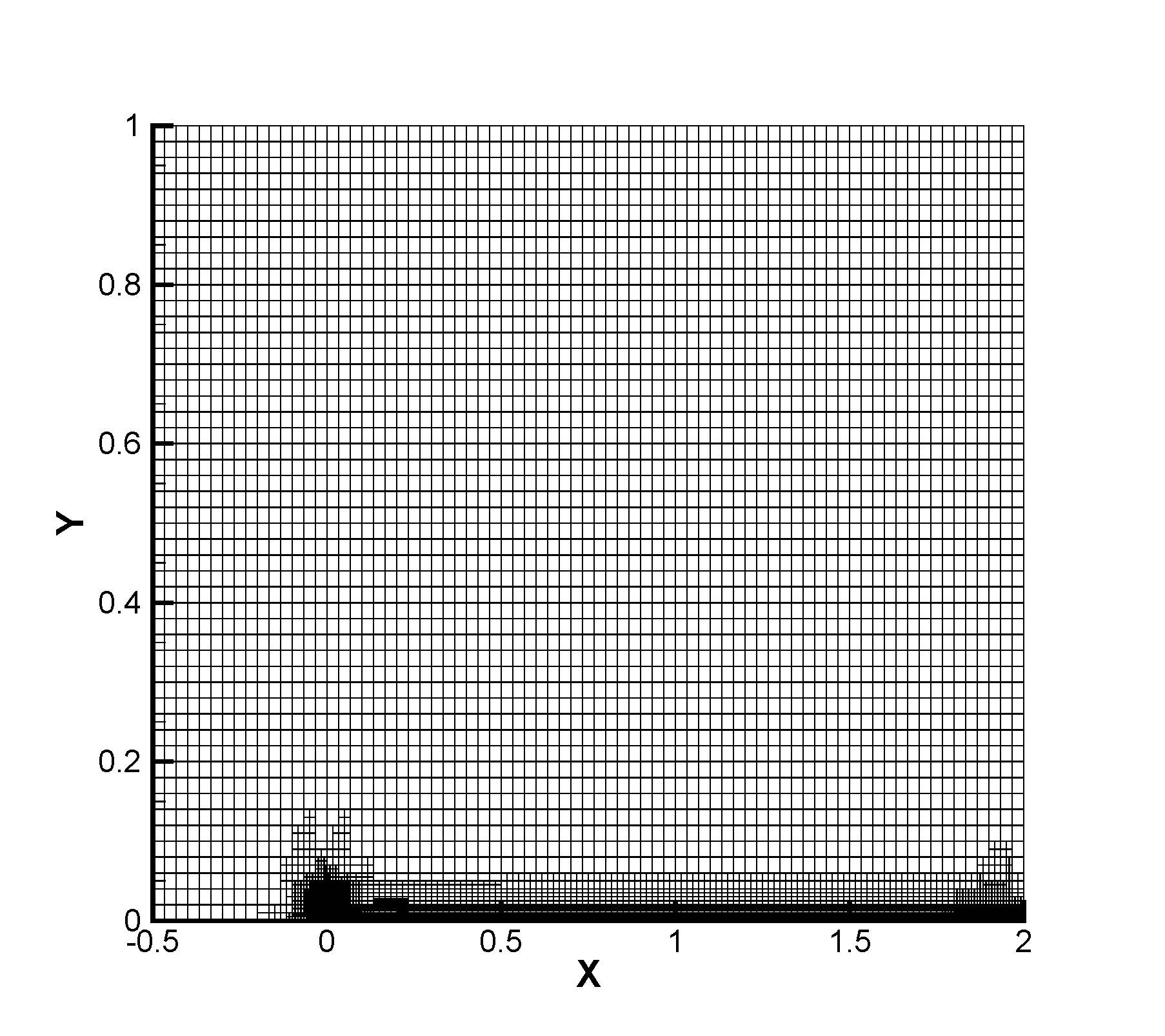}}
  \subfigure[]{\label{fig:Fig15b}\includegraphics[width=0.45\textwidth]{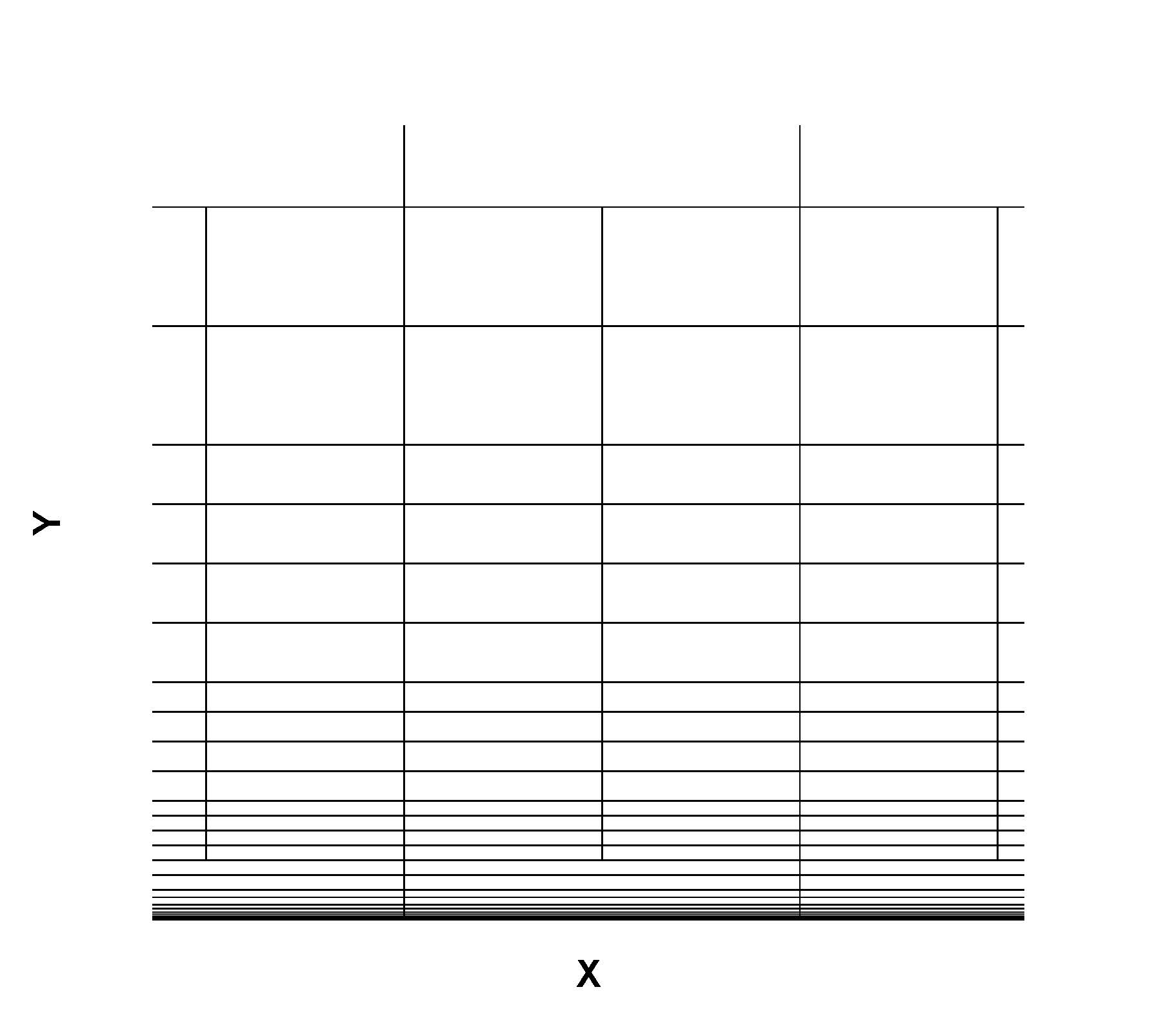}}
  \caption{The hybrid mesh for incompressible turbulent flow over a flat plate. \\(a) Hybrid mesh, (b) Mesh near wall.}
  \label{fig:Fig15}
\end{figure}

\begin{figure}
  \centering
  \subfigure[]{\label{fig:Fig16a}\includegraphics[width=0.45\textwidth]{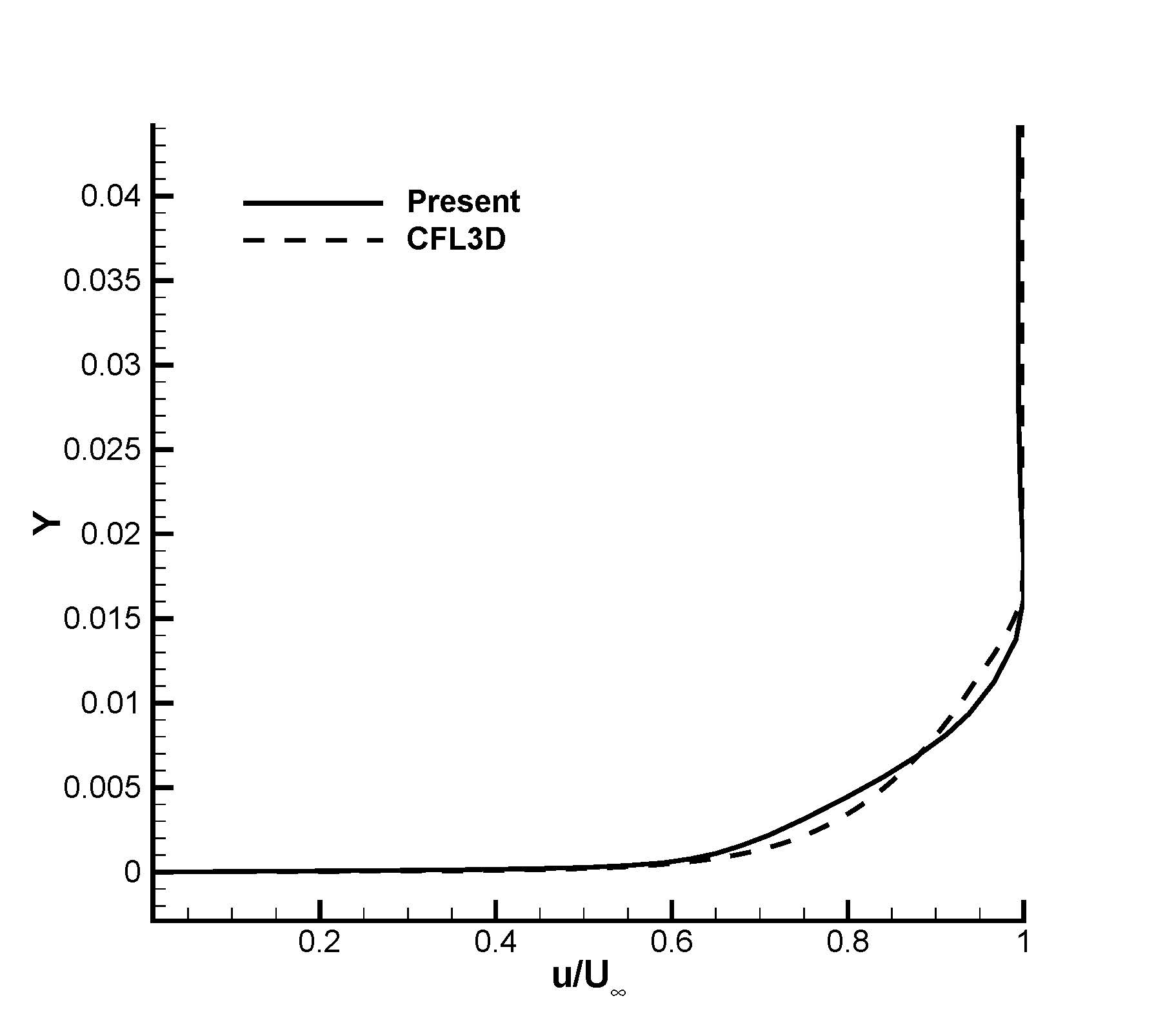}}
  \subfigure[]{\label{fig:Fig16b}\includegraphics[width=0.45\textwidth]{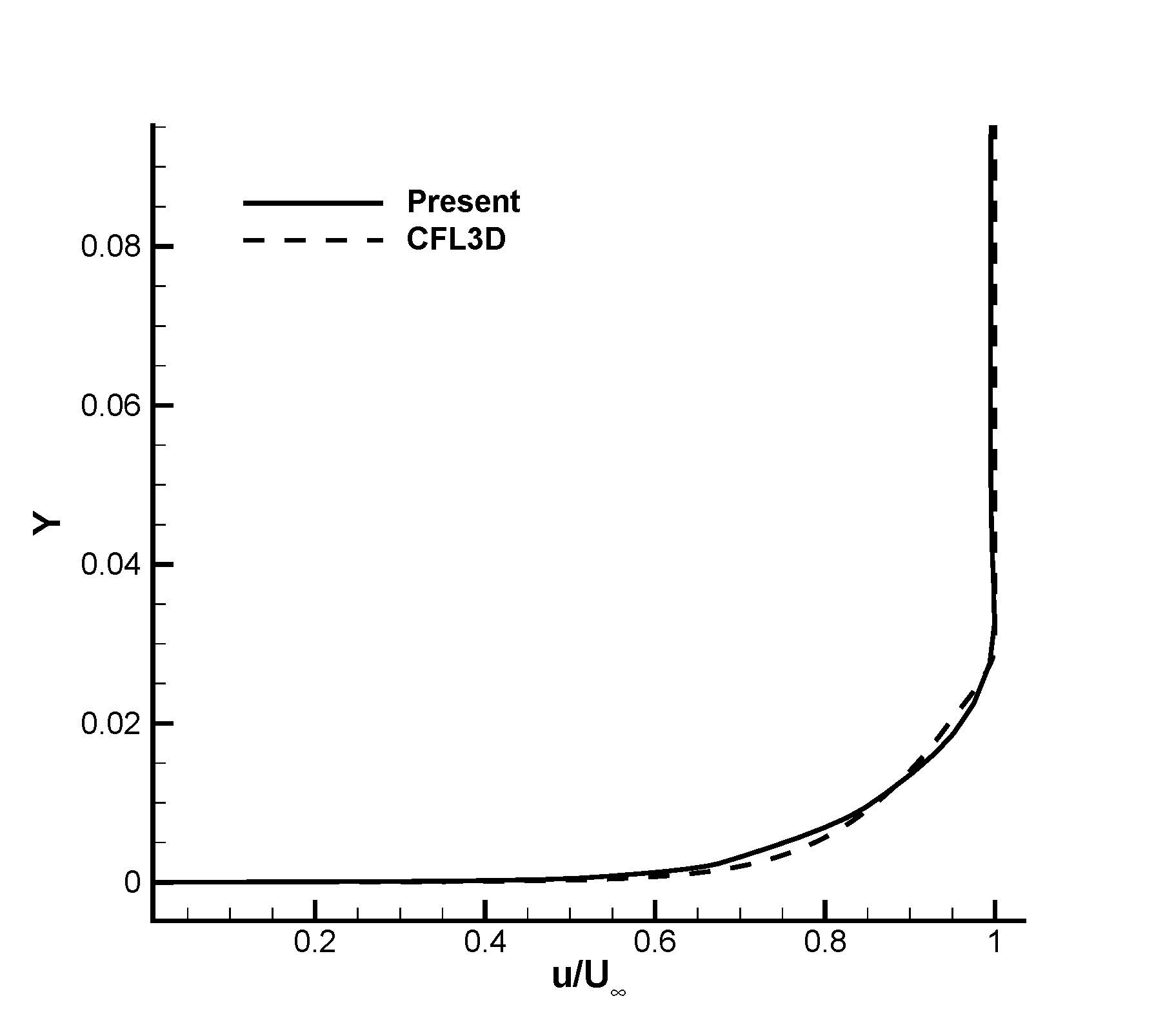}}
  \caption{Velocity profiles of incompressible turbulent flow over a flat plate. \\(a) $X = 0.97008$, (b) $X = 1.90334$.}
  \label{fig:Fig16}
\end{figure}

\begin{figure}
  \centering
  \includegraphics[width=0.5\textwidth]{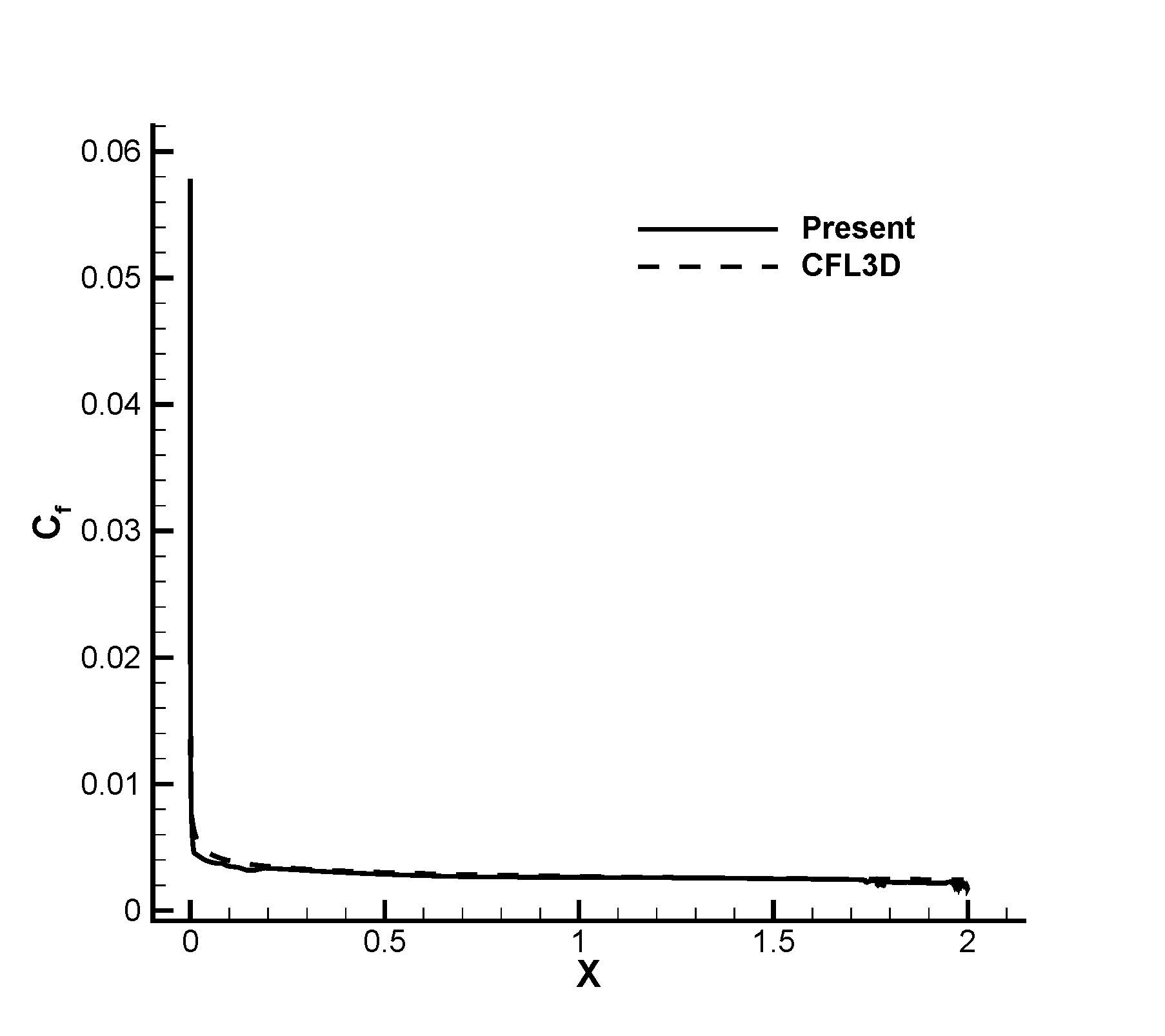}
  \caption{The skin friction coefficient over a incompressible turbulent flat plate.}
  \label{fig:Fig17}
\end{figure}

\begin{figure}
  \centering
  \includegraphics[width=0.5\textwidth]{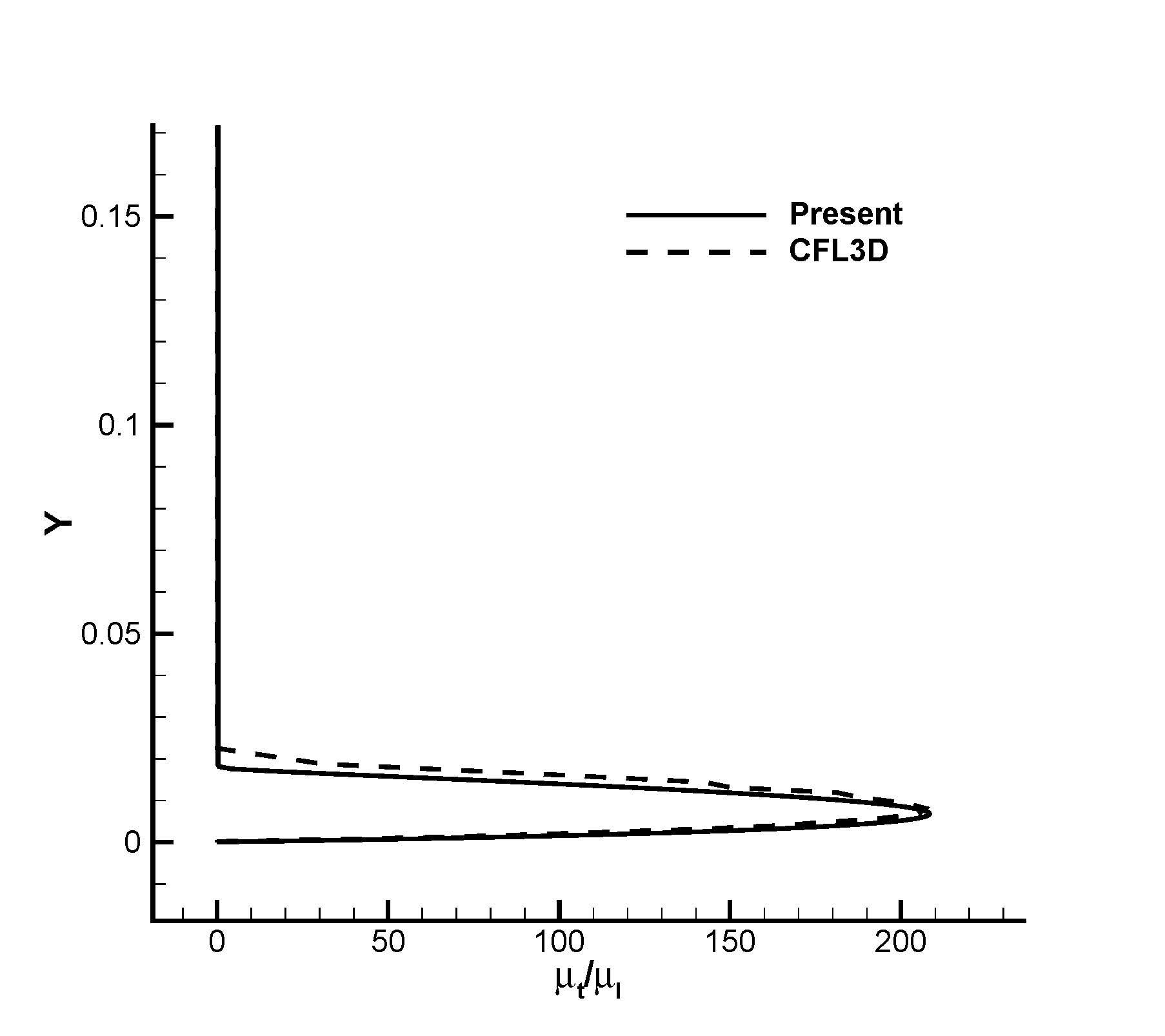}
  \caption{The nondimensional eddy viscosity of incompressible turbulent flow over a flat plate at position $X=0.97$}
  \label{fig:Fig18}
\end{figure}

\begin{figure}
  \centering
 \subfigure[]{\label{fig:Fig19a}\includegraphics[width=0.45\textwidth]{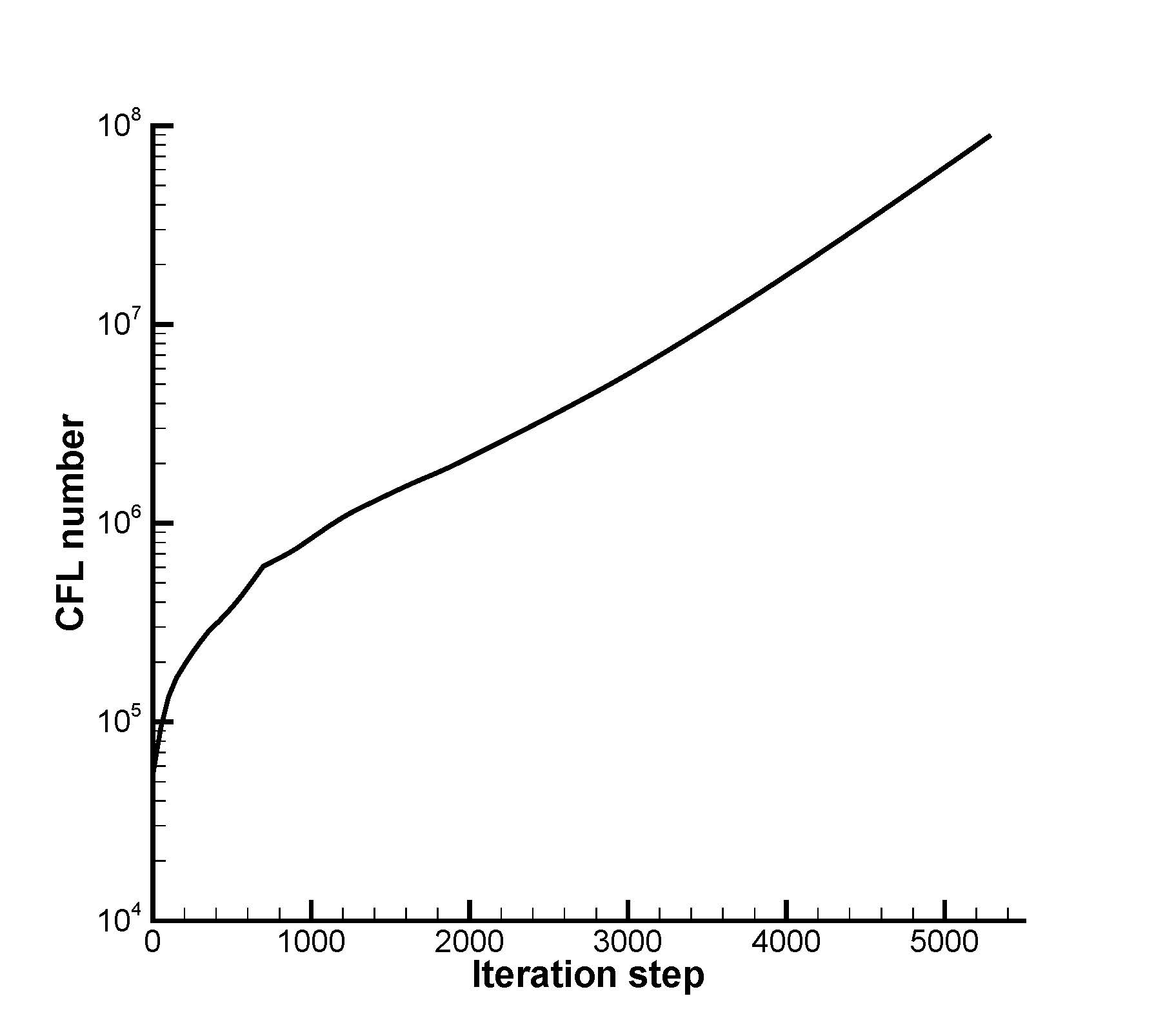}}
 \subfigure[]{\label{fig:Fig19b}\includegraphics[width=0.45\textwidth]{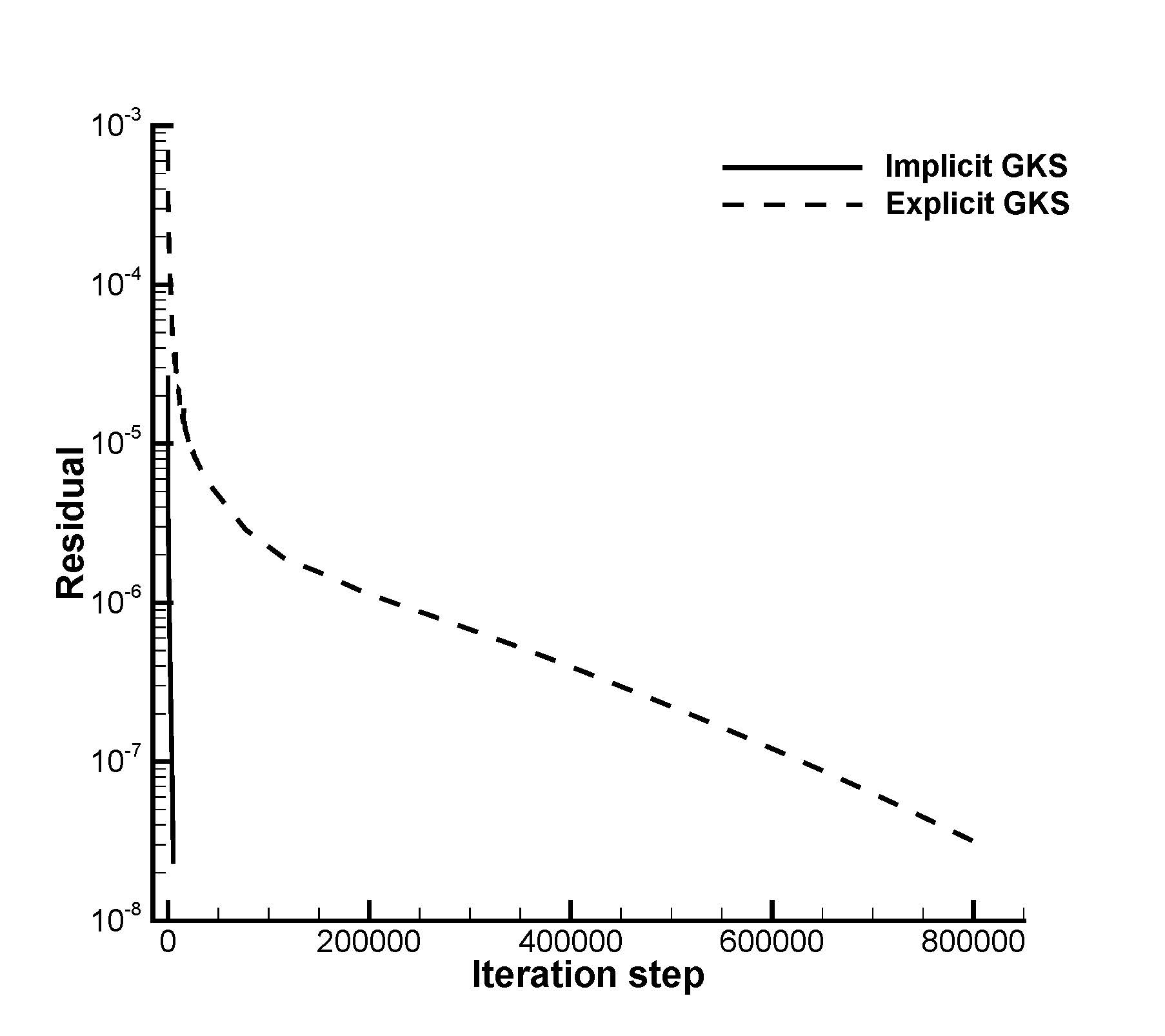}}
  \caption{The CFL number and residual curve of incompressible turbulent flow over a flat plate. \\(a) CFL number, (b) Residual.}
  \label{fig:Fig19}
\end{figure}

\begin{figure}
  \centering
  \includegraphics[width=0.5\textwidth]{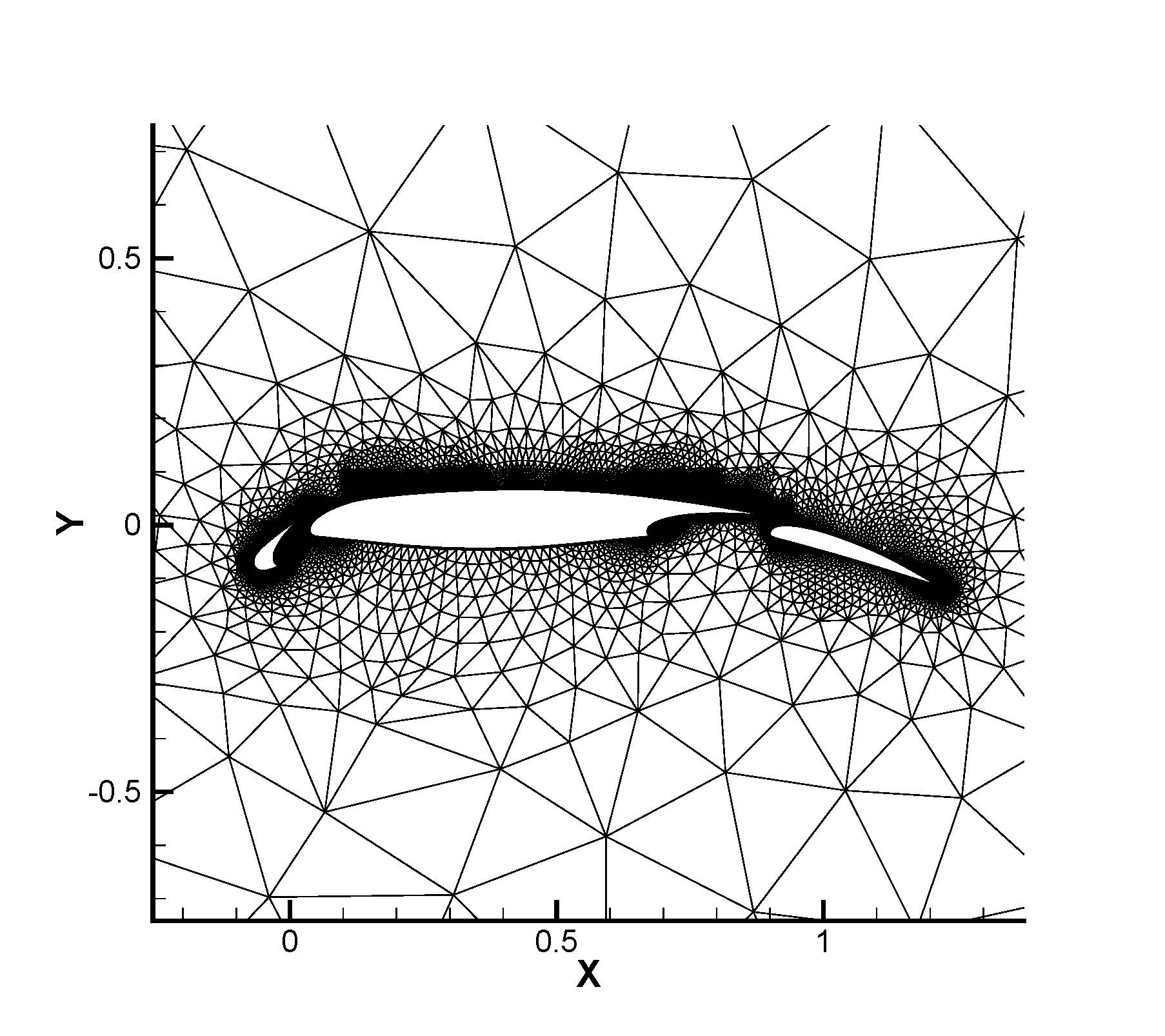}
  \caption{The unstructured mesh for flow around a NHLP multi-element airfoil.}
  \label{fig:Fig20}
\end{figure}

\begin{figure}
  \centering
  \subfigure[]{\label{fig:Fig21a}\includegraphics[width=0.45\textwidth]{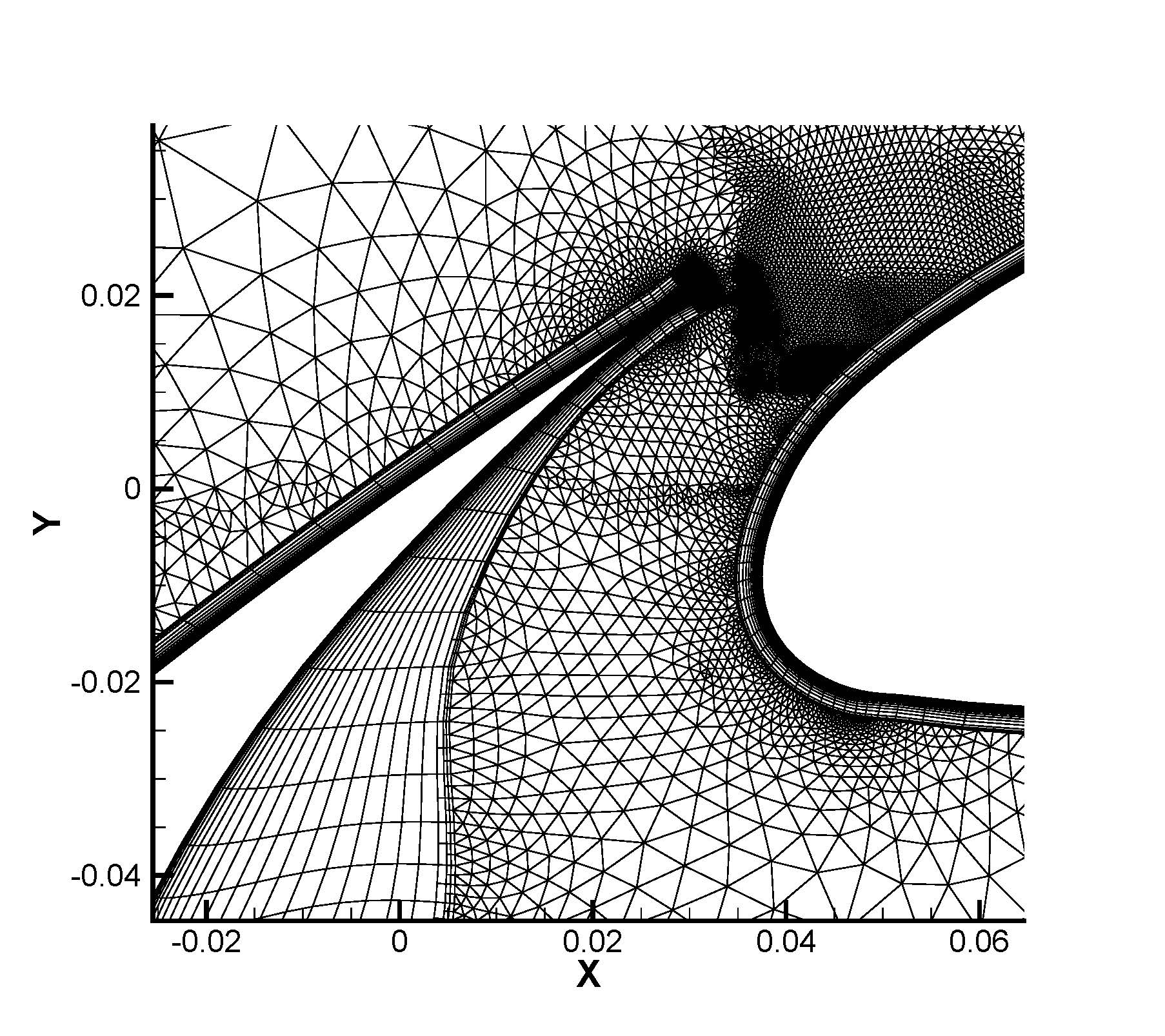}}
  \subfigure[]{\label{fig:Fig21b}\includegraphics[width=0.45\textwidth]{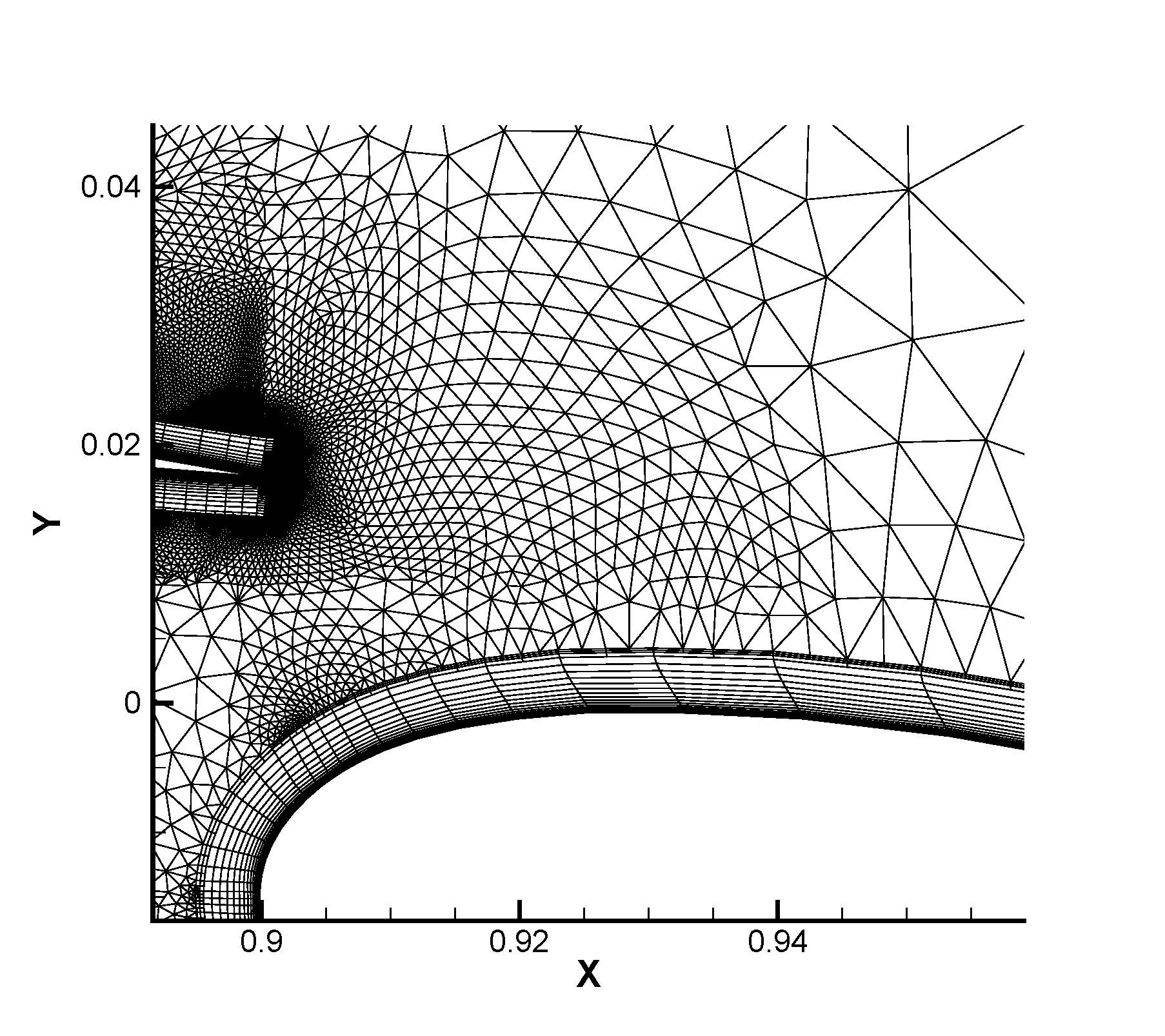}}
  \caption{The regional mesh for flow around a NHLP multi-element airfoil. \\(a) Between slat and airfoil, (b) Between airfoil and flap.}
  \label{fig:Fig21}
\end{figure}

\begin{figure}
  \centering
  \includegraphics[width=0.5\textwidth]{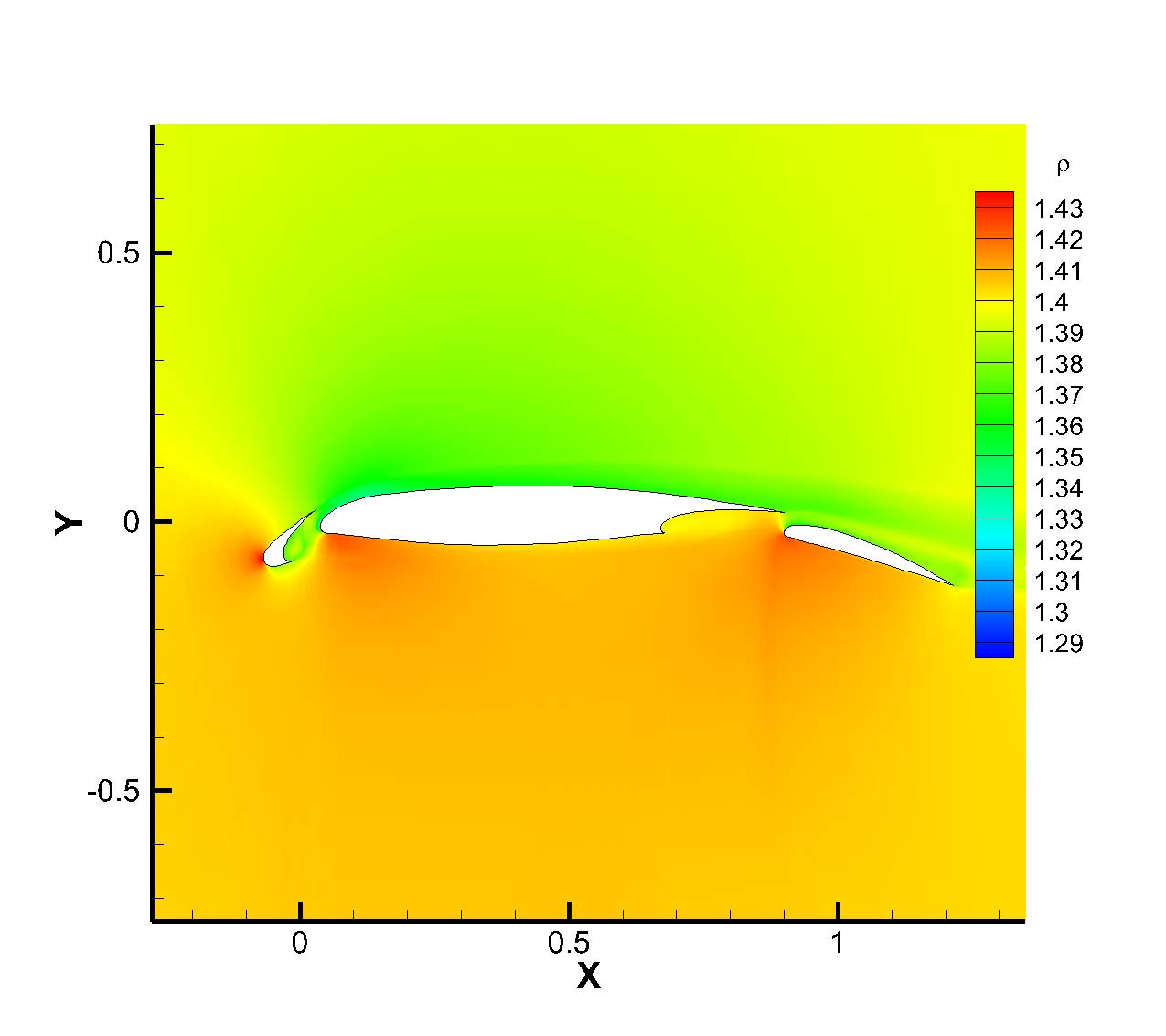}
  \caption{The density contour in flow around NHLP multi-element airfoil.}
  \label{fig:Fig22}
\end{figure}

\begin{figure}
  \centering
  \includegraphics[width=0.5\textwidth]{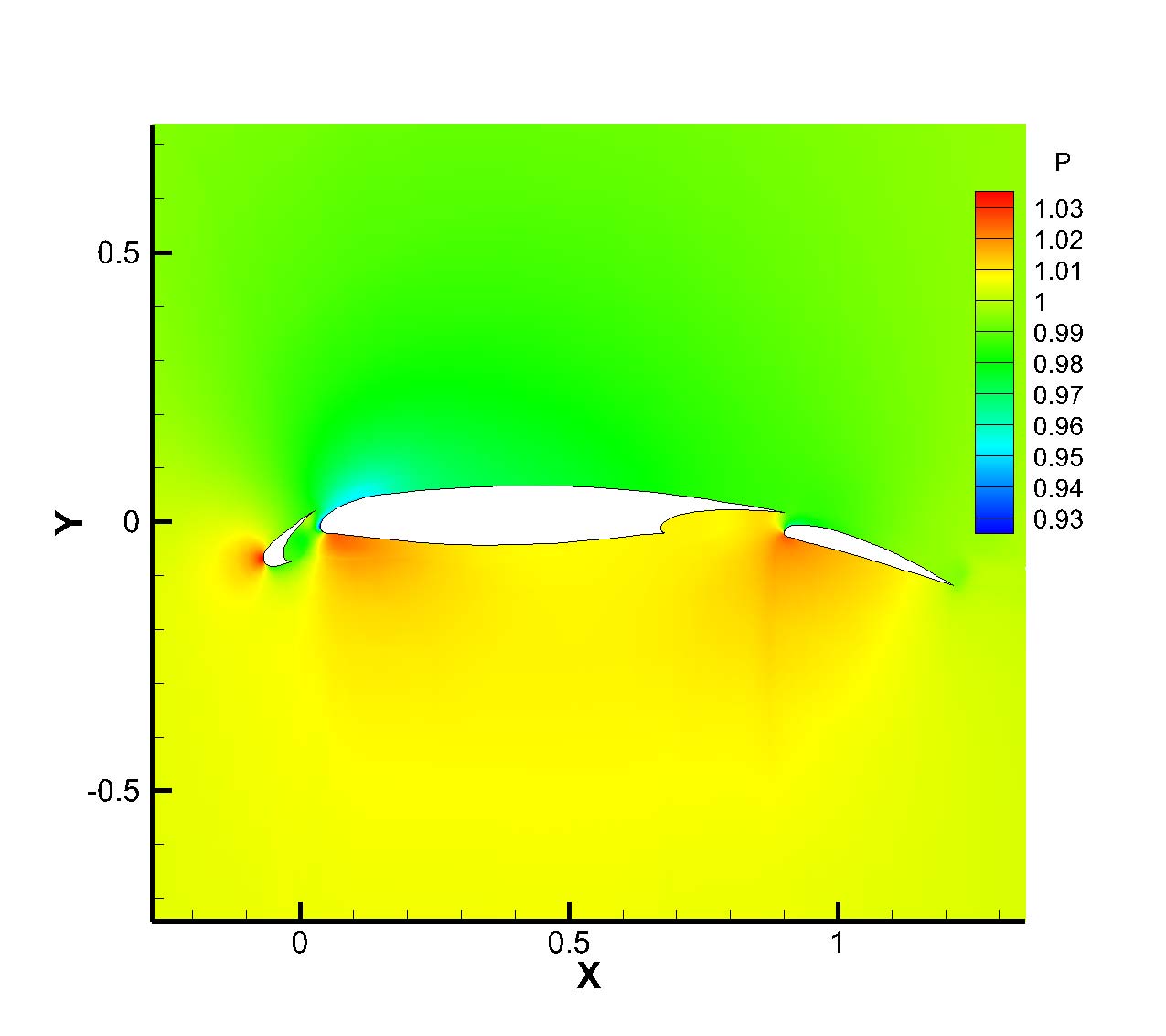}
  \caption{The pressure contour in flow around NHLP multi-element airfoil.}
  \label{fig:Fig23}
\end{figure}

\begin{figure}
  \centering
  \includegraphics[width=0.5\textwidth]{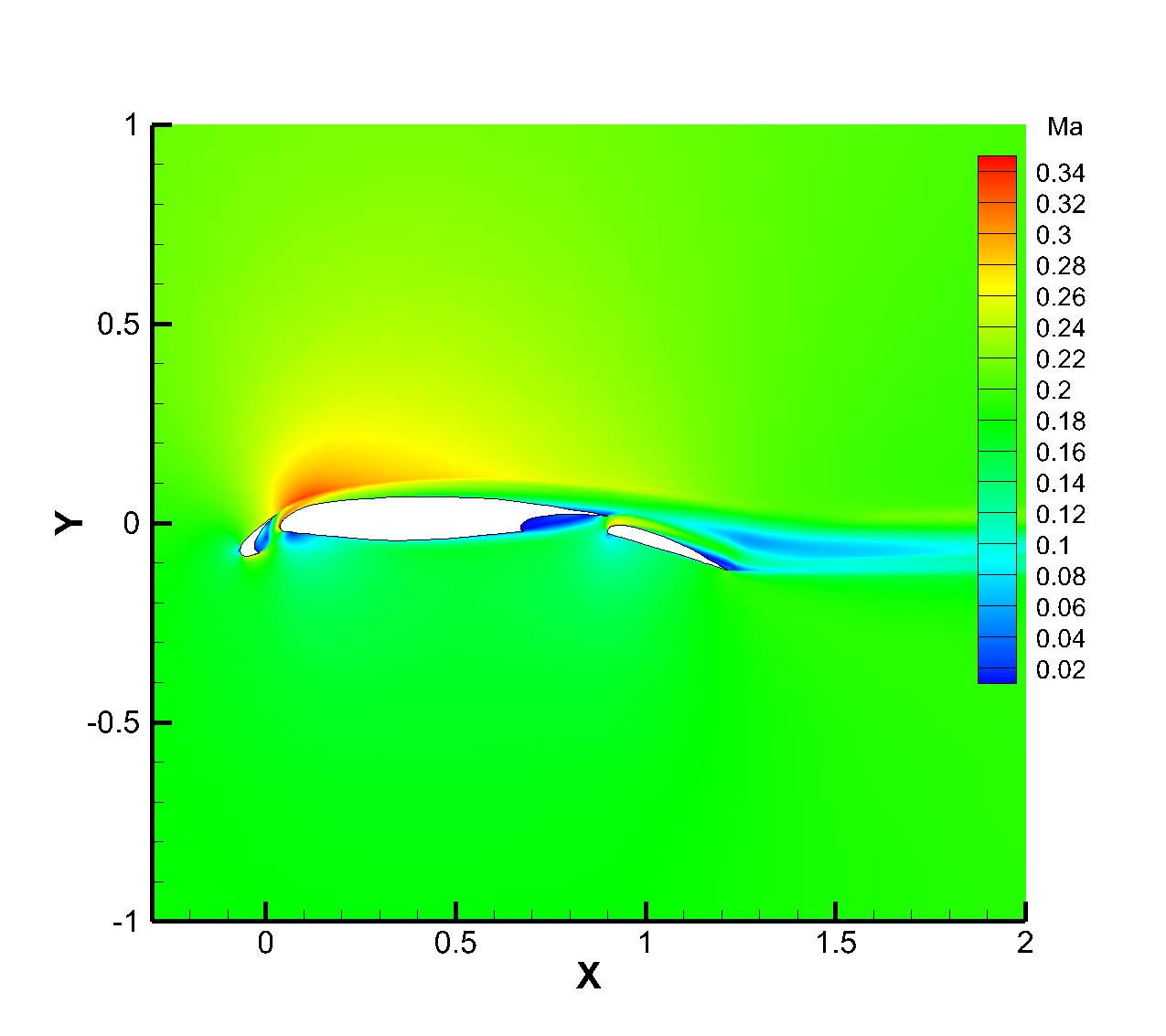}
  \caption{The Mach number contour in flow around NHLP multi-element airfoil.}
  \label{fig:Fig24}
\end{figure}

\begin{figure}
  \centering
  \includegraphics[width=0.5\textwidth]{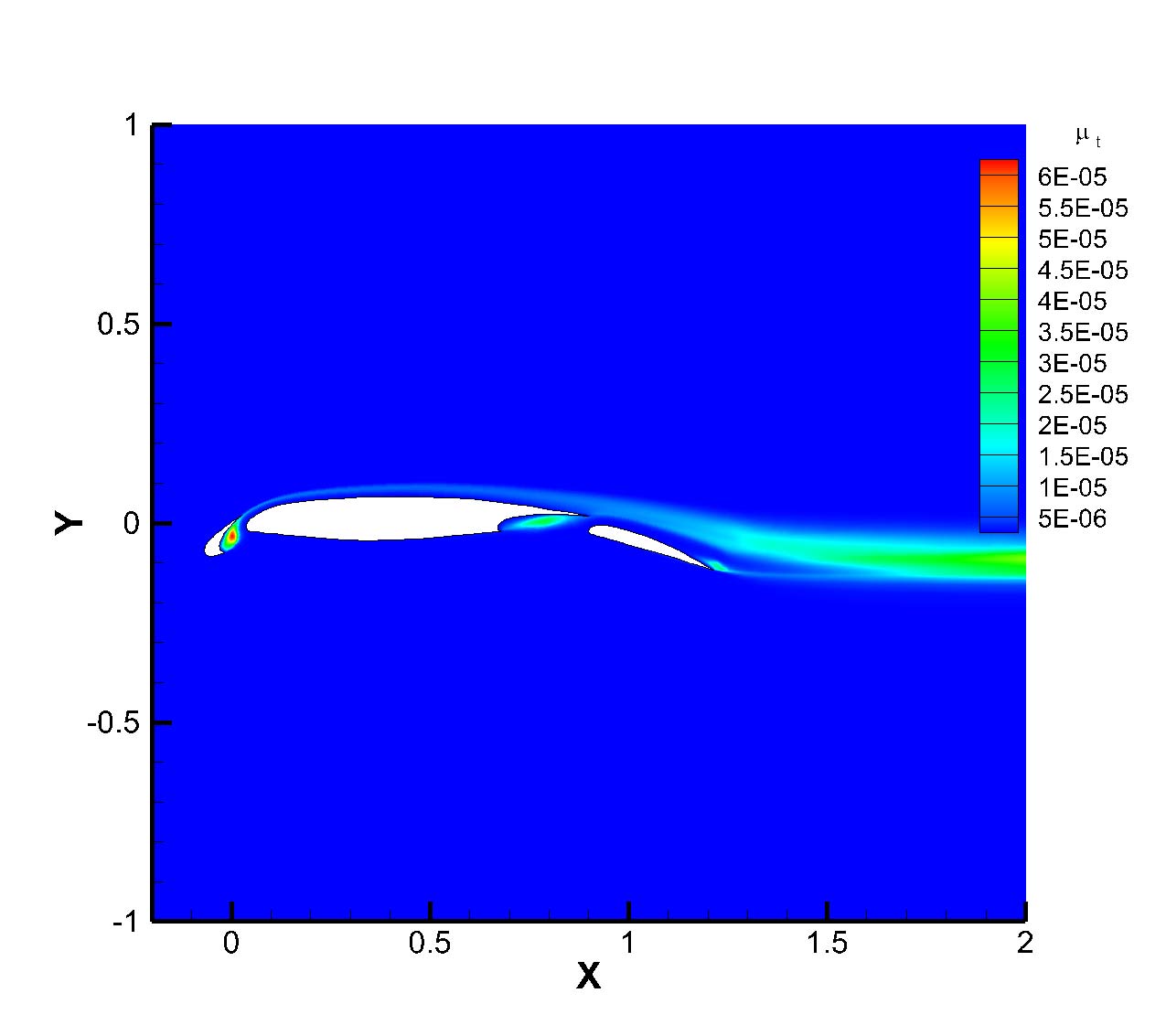}
  \caption{The eddy viscosity contour in flow around NHLP multi-element airfoil.}
  \label{fig:Fig25}
\end{figure}

\begin{figure}
  \centering
  \includegraphics[width=0.5\textwidth]{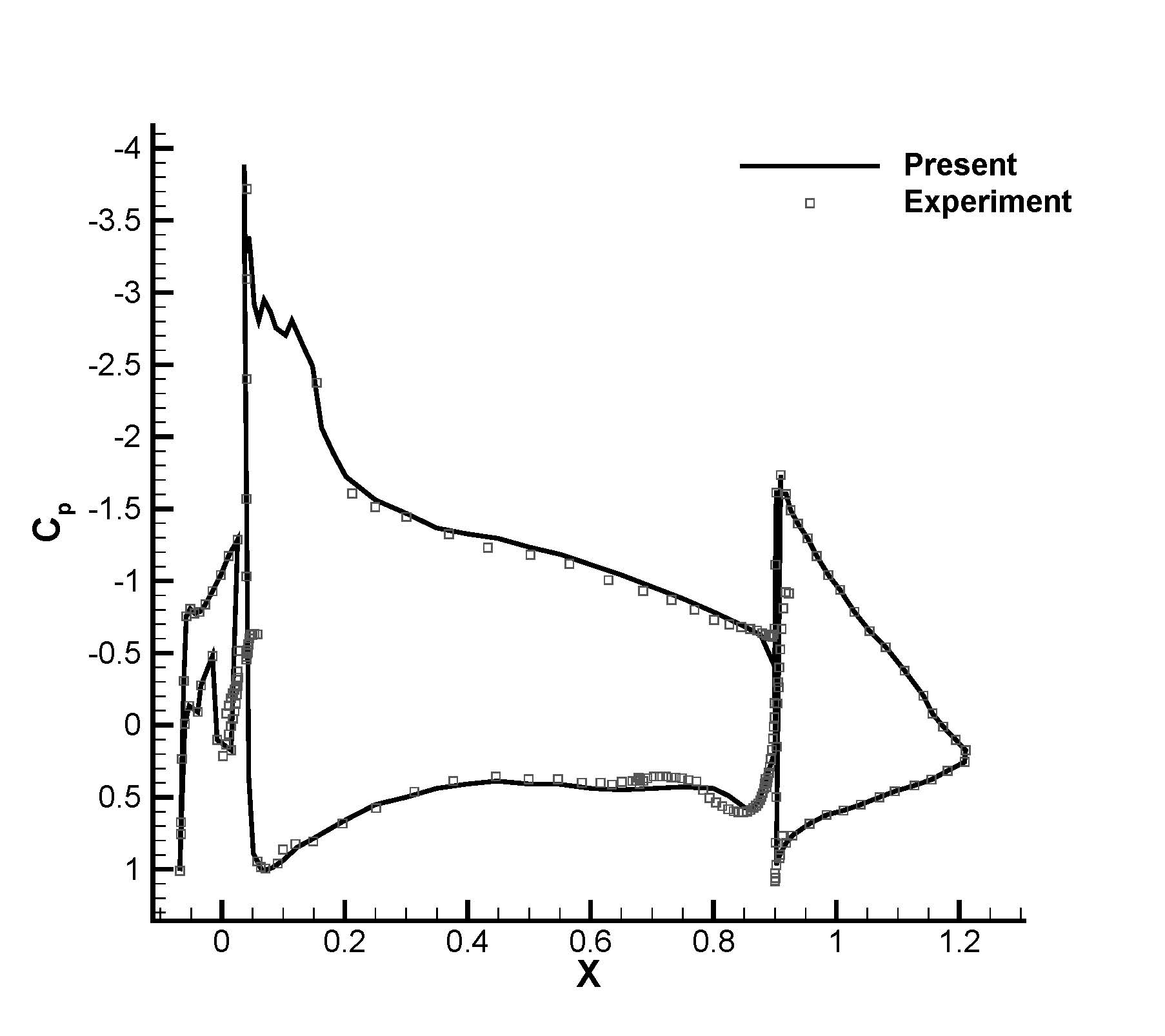}
  \caption{The pressure coefficient distribution over a NHLP multi-element airfoil.}
  \label{fig:Fig26}
\end{figure}

\begin{figure}
  \centering
 \subfigure[]{\label{fig:Fig27a}\includegraphics[width=0.45\textwidth]{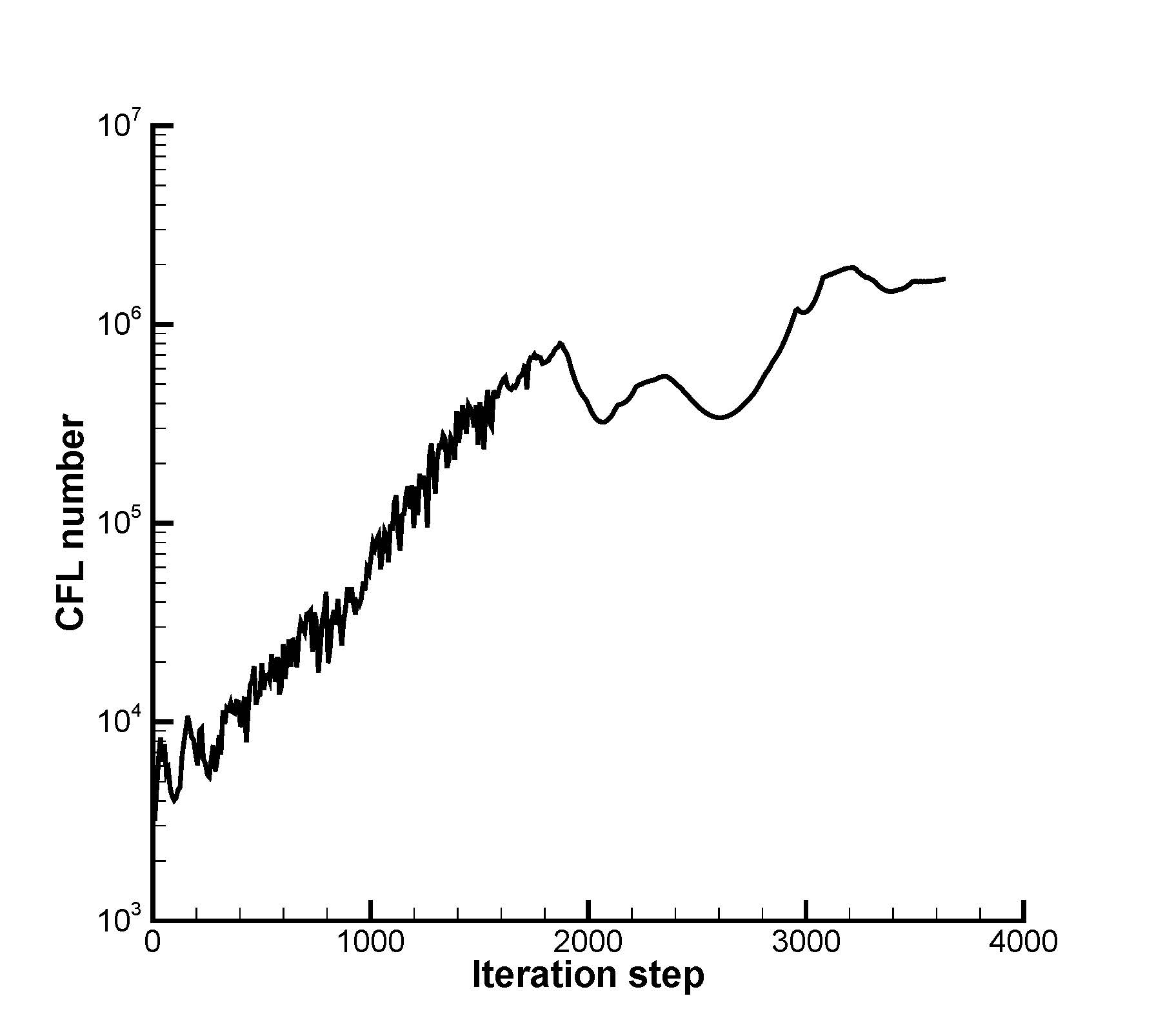}}
 \subfigure[]{\label{fig:Fig27b}\includegraphics[width=0.45\textwidth]{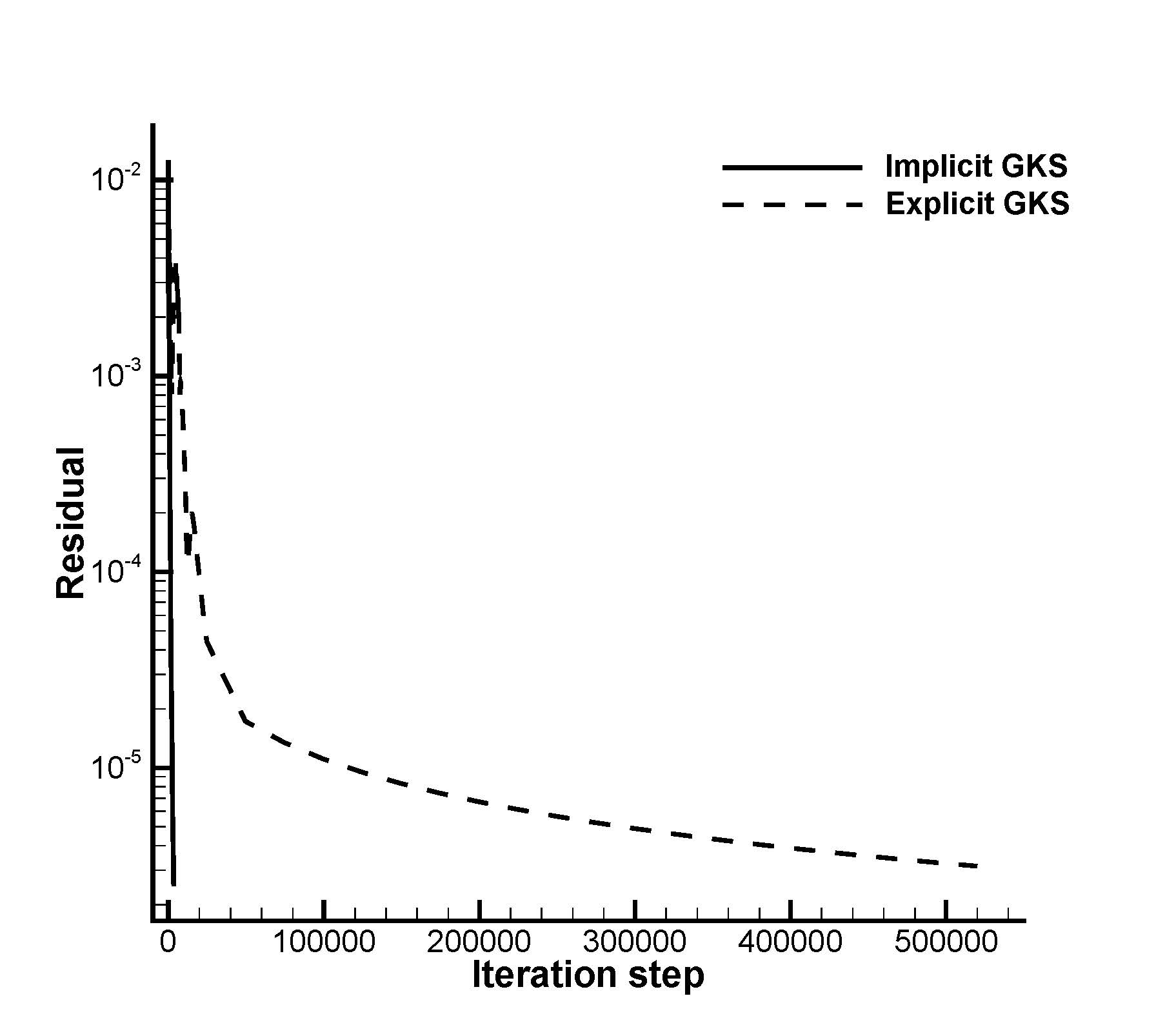}}
  \caption{The CFL number and residual curve in NHLP airfoil flow. \\(a) CFL number, (b) Residual.}
  \label{fig:Fig27}
\end{figure}

\end{document}